\documentclass[epj]{svjour-pp}
\usepackage{graphics}
\graphicspath{ {figures/} }
\usepackage{amssymb}
\usepackage{float}
\usepackage{chemmacros} 
\usepackage{chemfig} 
\usepackage[locale=US]{siunitx}
\DeclareSIUnit[number-unit-product = ]\percent{\char`\%}
\usepackage{fixltx2e}
\usepackage{color, soul}
\usepackage{url}
\reversemarginpar
\usepackage[para,online,flushleft]{threeparttable}
\usepackage{ulem}
\usepackage{upgreek}
\usepackage{placeins}
\usepackage{siunitx}

\usepackage{tikz}
\usetikzlibrary{decorations.pathmorphing}
\usetikzlibrary{calc}
\usepackage{gensymb}
\usepackage{subfigure} 
\usepackage{dblfloatfix}    
\usepackage{multirow}
\usepackage{cellspace}		
\usepackage[percent]{overpic} 

\newcommand\FEBIAD{VADIS}

\DeclareSIUnit\particle{p\hspace{-0.03cm}}

\begin{document}
\titlerunning{Extraction of refractory elements as carbonyl complex ions at CERN-ISOLDE}
\title{A concept for the extraction of the most refractory elements at CERN-ISOLDE as carbonyl complex ions}

\author{
	J. Ballof \inst{1,2} 
	\thanks{\email{jochen.ballof@cern.ch}}
	\and
	K. Chrysalidis 	\inst{1} 
	\and
	Ch.E. D\"ullmann\inst{2,3,4}
    \and
    V. Fedosseev \inst{1}
    \and
    E. Granados \inst{1}
    \and
    D. Leimbach\inst{1,5}
    \and
    B.A. Marsh\inst{1}
    \and
    J.P. Ramos \inst{1} 
    \thanks{Present address: SCK CEN, Boeretang 200, 2400 Mol, Belgium}
    \and
    A. Ringvall-Moberg \inst{1,7} 
    \and
    S. Rothe \inst{1} 
	\and
	T. Stora \inst{1}
    \thanks{\email{thierry.stora@cern.ch}}    	
    \and
    S.G. Wilkins \inst{1,8}
    \and
    A. Yakushev\inst{3,4}
}	
	

%
%
\institute{
	CERN, Systems Department, 1211 Geneva 23, Switzerland 
	\and
	Johannes Gutenberg - Universit\"at Mainz, Department for Chemistry, TRIGA-site, Fritz-Strassmann-Weg 2, 55128 Mainz, Germany
	\and
	GSI Helmholtzzentrum f\"ur Schwerionenforschung, 64291, Darmstadt, Germany
	\and
	Helmholtz-Institut Mainz, 55099 Mainz, Germany
	\and
	Johannes Gutenberg - Universit\"at Mainz, Department for Physics, Staudingerweg 7, 55128 Mainz, Germany
	\and
	{\'Ecole Polytechnique F\'ed\'erale de Lausanne (EPFL), Laboratory of Powder Technology, CH-1015, Switzerland}
	\and
	University of Gothenburg, Department of Physics, Gothenburg, Sweden
	\and
	University of Manchester, School of Physics and Astronomy, Manchester M13 9PL, United Kingdom
}
%

%
%
\abstract{We introduce a novel thick-target concept tailored to the extraction of refractory 4d and 5d transition metal radionuclides of molybdenum, technetium, ruthenium and tungsten for radioactive ion beam production. Despite the more than 60-year old history of thick-target ISOL mass-separation facilities like ISOLDE, the extraction of the most refractory elements as radioactive ion beam has so far not been successful. In ordinary thick ISOL targets, their radioisotopes produced in the target are stopped within the condensed target material and have to diffuse through a solid material. Here, we present a concept which overcomes limitations associated with this method. We exploit the recoil momentum of nuclear reaction products for their release from the solild target material. They are thermalized in a carbon monoxide-containing atmosphere, in which volatile carbonyl complexes form readily at ambient temperature and pressure. This compound serves as volatile carrier for transport to the ion source. Excess carbon monoxide is removed by cryogenic gas separation to enable low pressures in the source region, in which the species are ionized and hence made available for radioactive ion beam formation. The setup is operated in batch mode, with the aim to extract isotopes having half-lives of at least several seconds. We report parameter studies of the key processes of the method, which validate this concept and which define the parameters for the setup. This would allow for the first time the extraction of radioactive molybdenum, tungsten and several other transition metals at thick-target ISOL facilities.
} 
\maketitle
%
%
%

\section{Introduction}
\label{sec:intro}

The Isotope mass Separation OnLine (ISOL) technique was used already in 1951 to extract radioactive krypton isotopes from a 10 kg uranium oxide target, which was placed between the coils of a cyclotron magnet and irradiated by neutrons. The direct connection of the target to an ion source allowed simultaneous production and extraction of volatile species, and is today considered as the birth of the ISOL technique \cite{ISOLbirth1,ISOLbirth2}. Modern ISOL targets use less material (see ref. \cite{Ramos2019} for a recent review), but their areal densities are typically above several $\mathrm{g/cm^2}$. The thick target is both a blessing and a curse: the number of radioactive atoms produced inside the target scales with its thickness, translating into high yields for volatile elements like mercury. At CERN-ISOLDE, which is supplied by 1.4 GeV protons from the Proton Synchrotron Booster (PSB), extractable yields for \textsuperscript{197}Hg from a molten lead-bismuth target have been measured to be as high as \num{5e9}~ions  per $\mathrm{\upmu C}$ of protons, while the mean proton current for molten targets can reach up to $\mathrm{1.5\; \upmu A}$ \cite{LETTRY1997170,Catherall2017,Zanini2005}.
\begin{figure}
\resizebox{1\linewidth}{!}{%
  \includegraphics{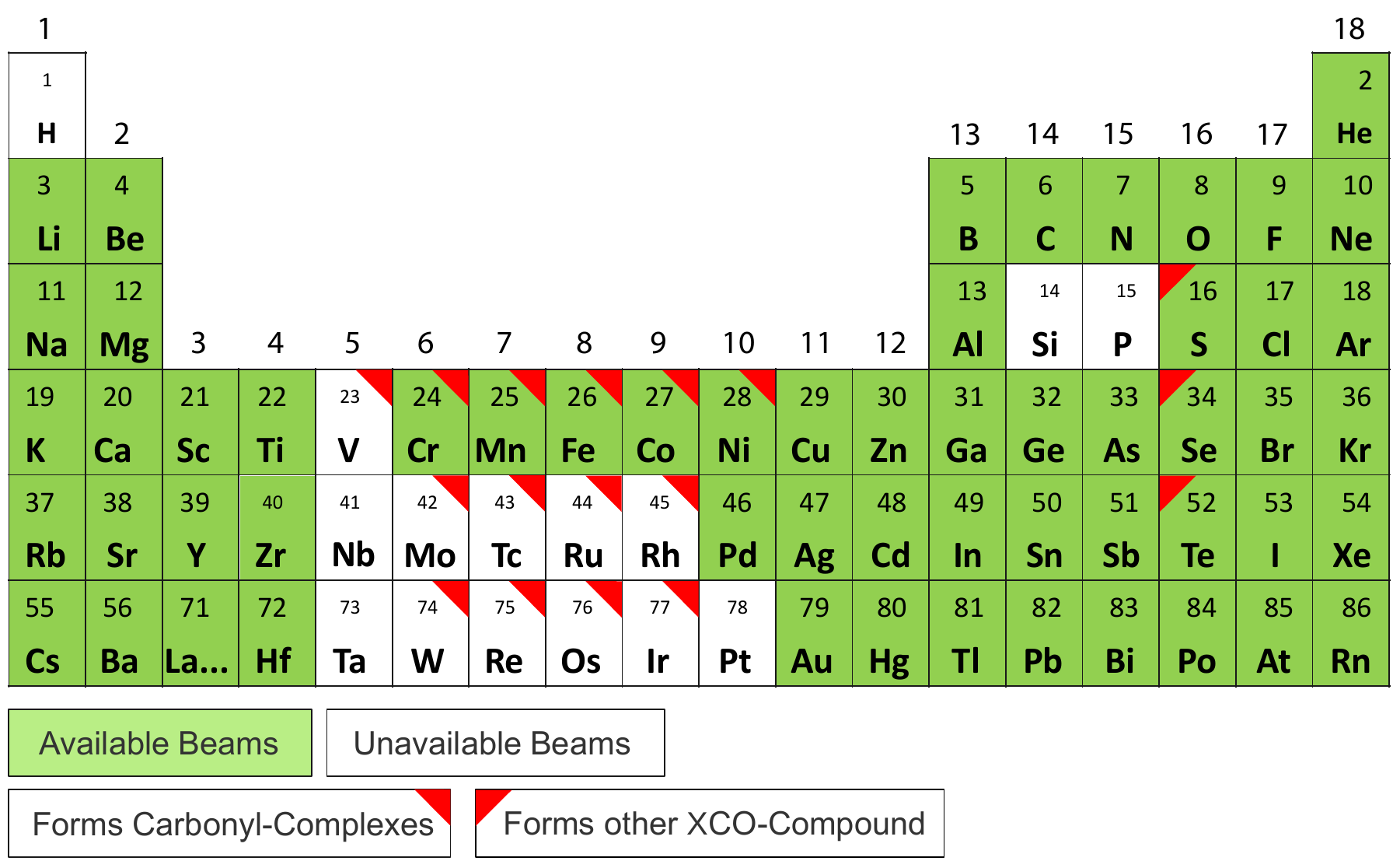}
}
\caption{Periodic table of elements showing available and not yet available thick-target ISOLDE-beams of short-lived isotopes ($T_{1/2} < \SI{1}{\hour}$), in comparison with the elements forming carbonyl compounds  \cite{elschenbroich2009organometallchemie,YieldDB,Liang1982}. S, Se and Te form compounds of type XCO (X= S, Se, Te), analogous to $\mathrm{CO_2}$. Only the first six rows of the periodic table rows are shown.}
\label{fig:pse}
\end{figure}

\begin{figure*} [b]
\resizebox{1\textwidth}{!}{%
  \includegraphics{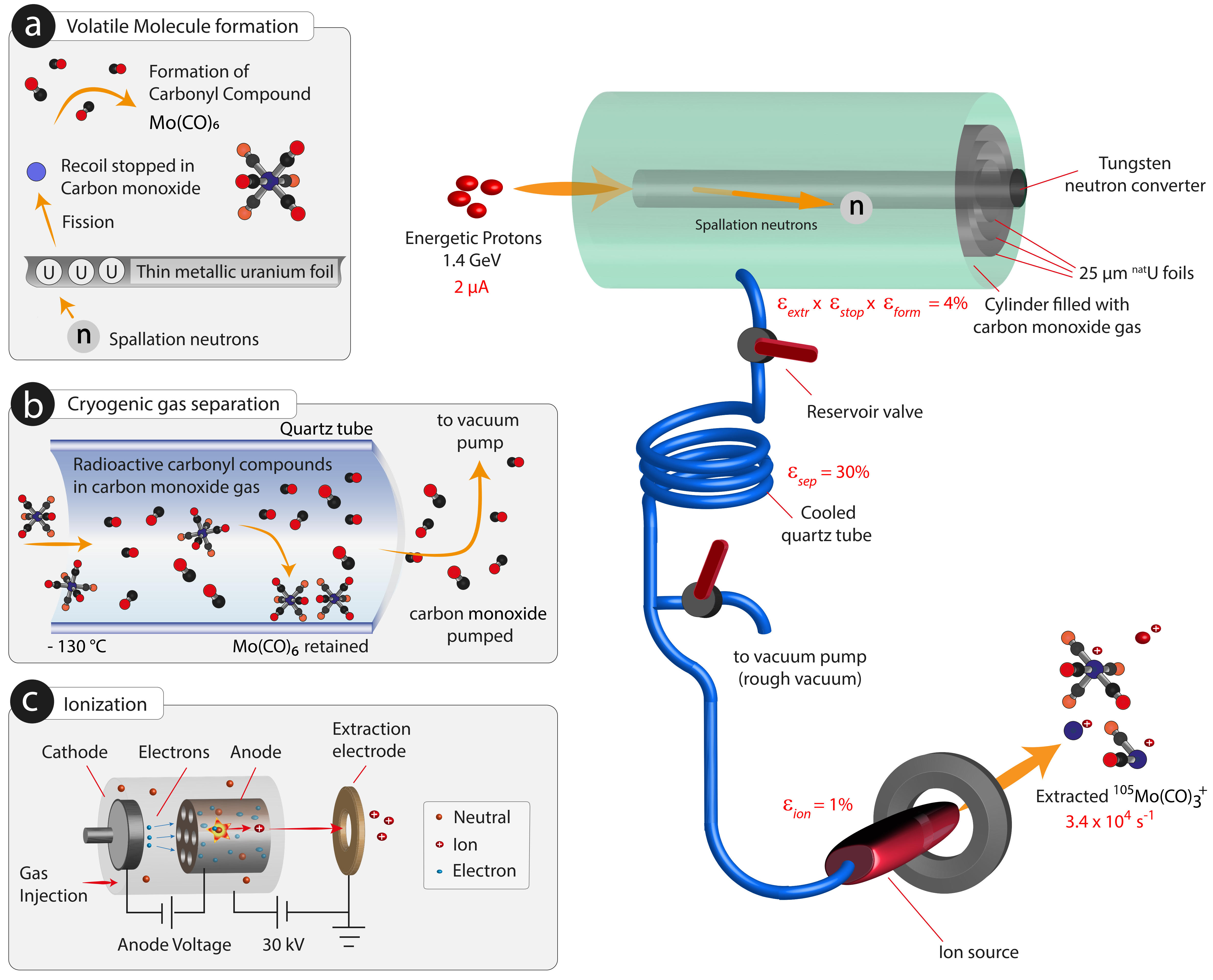}
}
\caption{Target concept for the extraction of refractory transition metals from a metallic foil target, shown exemplarily for the extraction of \textsuperscript{105}Mo from uranium foils. Highly energetic protons impinge on a tungsten neutron converter and produce spallation neutrons. (a) The neutrons induce fission in the uranium foils and fission recoils emerge from the foil and are thermalized in carbon monoxide. Upon thermalization molybdenum hexacarbonyl complexes form readily. (b) The excess carbon monoxide is removed by cryogenic gas separation and the carbonyl complexes are fed into the ion source. (c) Electron impact ionization is proposed to ionize molybdenum hexacarbonyl. Electrons emerging from a cathode are accelerated through a grid into a cylindrical anode where ionization takes place in collisions between neutral molecules and electrons. The ion source is kept at a potential of \SI{+30}{\kilo\volt} and ions are extracted through the grounded extraction electrode. Further details are given in the text.}
\label{fig:productionschema}
\end{figure*}
Issues arise, if i) the desired element is either chemically reactive and froms strong bonds to the target or to structural materials of the target and ion source system, ii) the diffusion of the element in the target material is slow, or iii) if the effusion through open space from the target to the ion source is hindered due to long sticking times at each wall encounter. The mean sojourn time $\tau$ for a single wall collision can be estimated by the Frenkel equation \cite{Frenkel,Frenkel1924},
\begin{equation}
\label{eq:frenkel}
\tau = \tau_0 \; \mathrm{e}^{- \Delta H_{\mathrm{ads}} / (\mathrm{R}T)},
\end{equation}
where R is the universal gas constant, $\tau_0$ the period of vibration perpendicular to the surface in the adsorbed state and $\Delta H_{\mathrm{ads}}$ the enthalpy of  adsorption, which is a measure of the interaction strength between a single atom, molecule or ion and a surface. Hence, $\tau_0$ is a property of the adsorbent only, while  $\Delta H_{\mathrm{ads}}$ depends on adsorbent and adsorbate. The latter can often be correlated to the macroscopic sublimation enthalpy of the adsorbent, characterizing the interaction between atoms or molecules of the same kind, which in turn is related to the vapour pressure of the compound. Hence, the vapour pressure is a measure for the mobility of a species during effusion processes for a given surface \cite{SuperheavyChemistry}. As can be seen from Eq.~\ref{eq:frenkel}, the mean sojourn time depends exponentially on the temperature. Elements suffering from low vapour pressure and high boiling point are called refractory, and their extraction as ISOL beam provides a substential challenge. Volatile carrier molecules need to be formed, to enable transport of these elements to the ion source. From there they are either extracted as a molecular beam even as elemental ion beam, due to dissociation in the ion source. The process is called \textit{in-situ} volatilization, chemical evaporation or extraction as molecular sideband. Reviews by K\"oster \textit{et al.} give an excellent overview of the method and its limitations \cite{ImpossibleBeamskoester,progresskoester}. A recent development, not yet listed therein, is the successful extraction of the refractory and chemically reactive element boron as fluoride \cite{boronpaper}. 
The elements for which short-lived isotope beams ($T_{1/2} < \SI{1}{\hour}$) are available from thick-target ISOL facilities are shown in fig.~\ref{fig:pse}. 
The most volatile elements, like noble gases, can already be extracted at low target temperatures. As both the sticking times and the diffusion processes inside the target material depend exponentially on the temperature, it is possible to extract most of the elements with the ISOL technique, if the target is operated at elevated temperatures. Therefore, the ISOLDE target container is made of tantalum and is designed to be heated restively to a maximum temperature of ca. 2250 \degree C.

%

As illustrated in fig.~\ref{fig:pse}, many transition metals are not yet available as ISOL radioactive ion beam. This region of the periodic table contains the most refractory elements like molybdenum, tantalum, tungsten and rhenium. These are at the same time the construction materials of the typical target and ion source unit. So far, it was only possible to extract tantalum as TaF\textsubscript{5} \cite{Liang1982,Liang1985}, which is compatible with the structural materials of a FEBIAD ion source. 
Long-lived \textsuperscript{99m}Tc ($T_{1/2} = \SI{6}{\hour}$) could be extracted in elemental form from a tantalum carbide target equipped with a hot rhenium cavity by resonant laser ionisation. The online extraction of short-lived Tc isotopes is not reported \cite{triumfYieldDB}.
Suitable sidebands for the remaining 4d and 5d refractory metals have not yet been found. 

The requirements for volatile carriers are multifold. The compound should form easily upon reaction of a produced radioactive atom and a reaction partner, which is either present in the target and ion source system or introduced via injection of reactive gases. The latter can be either directly injected into the target or produced \textit{in-situ} by heating a small solid sample which is connected to the target via a tube. 

After formation of the isotope carrier compound, it has to travel from the target container via a transfer line to the ion source. On its way numerous wall encounters occur. Hence, the compound must be chemically inert towards the materials present in the target and ion source system. High temperatures are often beneficial for fast diffusion and volatile compound formation. However, it can be adverse to the desired chemical stability of the carrier molecule. This is especially the case, if decomposition is favored by thermodynamics, but could be inhibited kinetically, \textit{i.e.} by slower reaction rates at lower temperatures. Moreover, if decomposition is required for beam purification purposes, catalytic processes at low temperatures can be considered.

After eventually reaching the ion source, the compound needs to be ionized efficiently. 
For the extraction of the 4d and 5d transition metals of groups 6 to 9, which the present article is addressing, the extraction as fluoride or oxide has been discussed \cite{ImpossibleBeamskoester,Liang1982,Liang1985}. Both compound classes suffer from limited compatibility with the materials present in target and ion sources. This especially holds for tantalum, which forms strong bonds to oxygen and fluorine, thus decomposing the volatilized oxygen or fluor-containing molecules upon impact.

Another issue arises from the target material. Typically uranium carbide targets are used at ISOLDE to produce heavy elements, fission products and light neutron-rich fragments. The reductive environment and the presence of carbon in the material are adverse to oxidation reactions. The material structure of uranium oxide has only limited thermal stability and is prone to fast sintering, which increases the diffusion time. To follow a classical approach for the extraction of the refractory metal beams, the development of a new target material, ideally a target and transfer line free from metallic surfaces, and the integration of an efficient ECR (Electron Cyclotron Resonance) ion source would be required, due to catalytic decomposition of the volatile carrier on metallic surfaces and the given source temperatures. However, even after the implementation of all of these developments, the issue of slow diffusion in the target material remains. Hence, we propose a new target concept to bypass the limitations of a more classical high-temperature approach.


\section{The target concept}
\label{sec:concept}

To overcome operation at high temperatures, a novel target concept is proposed, allowing target operation at ambient temperature and pressure. The new concept aims at producing and extracting radioisotopes with half-lives of at least one to ten seconds in a batch-mode.

\begin{figure*}
    \centering
    \subfigure[Proton fluence]{\includegraphics[width= 0.55 \textwidth]{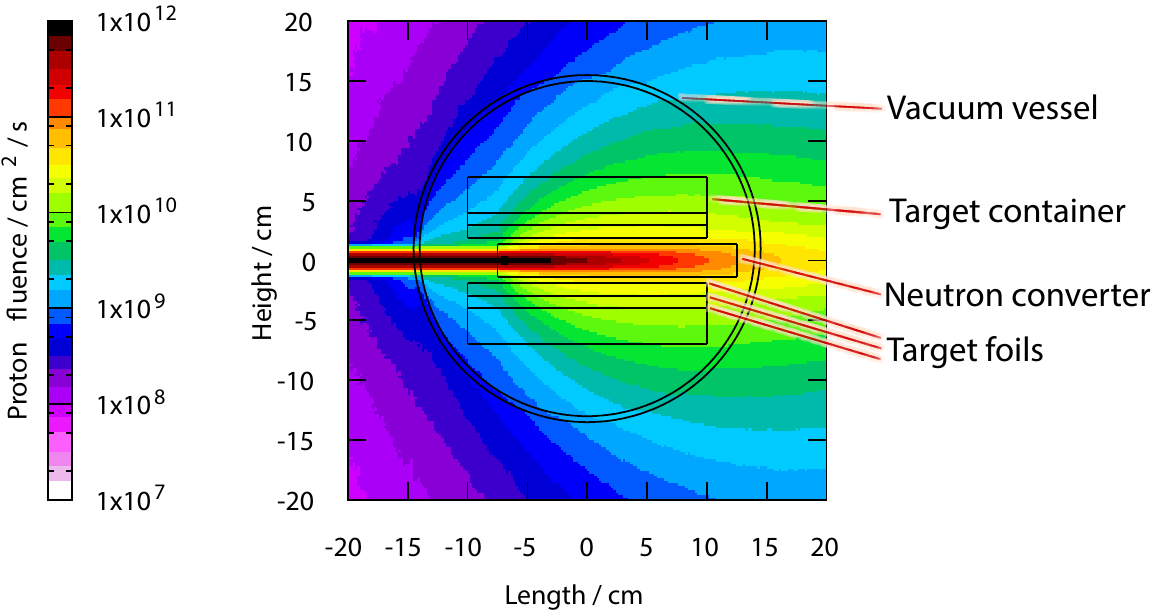}\label{sugfig:proton_fluence}}
    \hspace{0.2cm}
    \subfigure[Neutron fluence]{\includegraphics[width= 0.385 \textwidth]{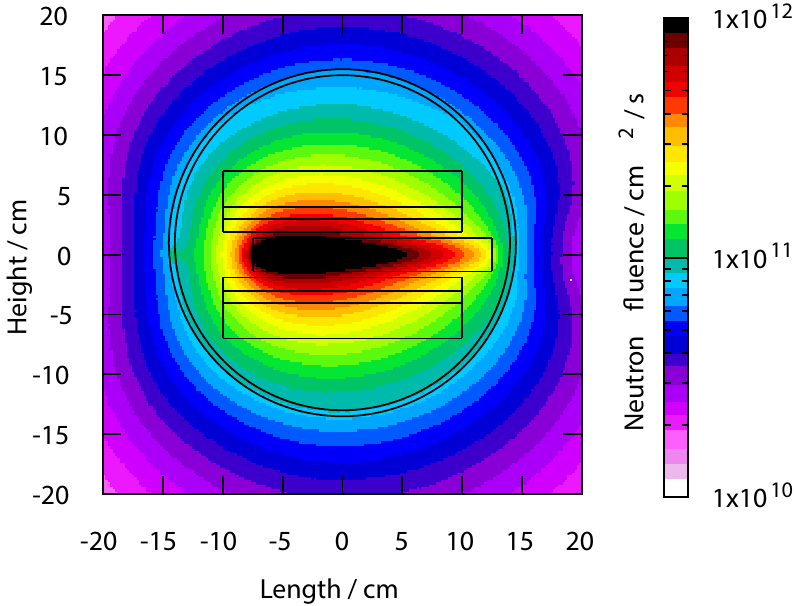} \label{sugfig:neutron_fluence}}
    \caption{(a) Proton and (b) neutron fluences at irradiation with $1\times 10^{13}$ protons per second obtained by FLUKA for the uranium foil geometry described in the text.  The proton beam arrives from the left on the neutron converter. A sketch of the target is shown in fig.~\ref{fig:productionschema}. }
 \label{fig:FLUKAfluence}
\end{figure*}

\begin{figure}[b]
    \centering
\resizebox{0.9 \linewidth}{!}{%
      \subfigure{\includegraphics[width=0.49\textwidth]{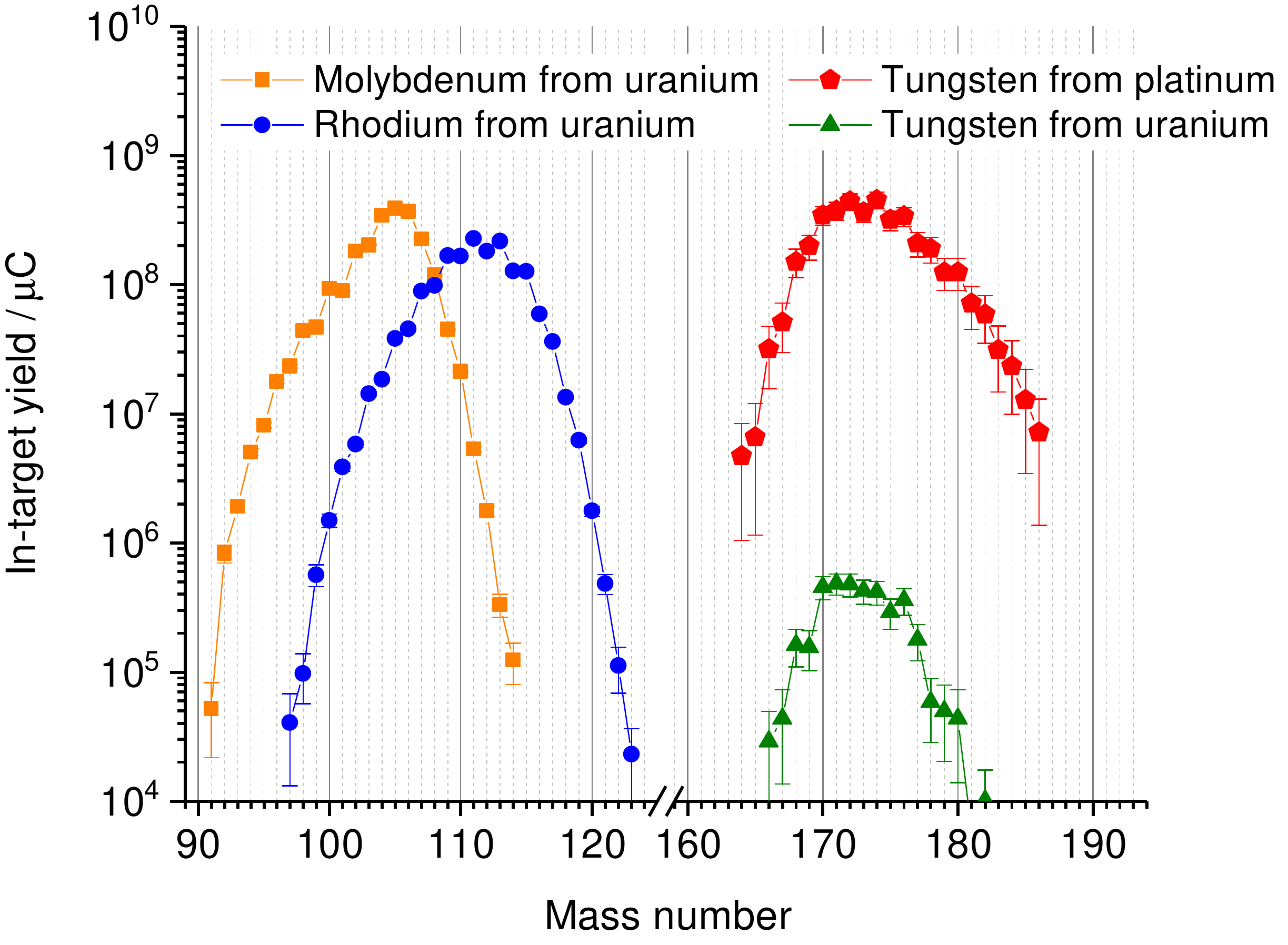}} 
    }
    \caption{In-target production for selected elements obtained by FLUKA. The error bars correspond to statistical errors of the simulation. The target geometry parameters are listed in Tab. \ref{tab:foilgeometries}.}
 \label{fig:FLUKAprod}
\end{figure}

The recoil momentum allows ejectils of nuclear reactions to propagate through and emerge from thin metallic foils. In the proposed concept, the recoils are thermalized in a reactive gas and form volatile compounds \textit{in-situ} (cf. fig.~\ref{fig:productionschema}). A compound class that appears well suited for \textit{in-situ} volatilization is that of metal carbonyl complexes. As can be seen in fig.~\ref{fig:pse}, nine out of fifteen transition metals, of which beams are not yet available, form volatile carbonyl compounds. Already in 1961, the extraction of molybdenum from uranium oxide by formation of a volatile carbonyl compound was demonstrated \cite{Baumgart1961}. Baumg\"artner and Reich\-hold irradiated a mixture of $\mathrm{U_3O_8}$ and $\mathrm{Cr(CO)_6}$ with neutrons and were able to extract molybdenum hexacarbonyl by sublimation. Later, nuclear reaction products where thermalized in a chromium hexacarbonyl catcher, from which they could be evaporated as carbonyl compound \cite{Wolf1973}. However, a catcher made of $\mathrm{Cr(CO)_6}$ is incompatible with an ion source as used in ISOL target units, due to its high volatility \cite{Baechmann1970}. Recently, it was shown by Even \textit{et al.} that volatile carbonyl complexes readily form at ambient temperature and pressure by thermalizing fission fragments of suitable elements in a carbon monoxide-containing atmosphere \cite{even2014situ,Even_2012}. Within our concept, the formation of the volatile compound is followed by removal of the reactive gas by cryogenic gas separation \cite{Katagiri2015}, as illustrated in fig.~\ref{fig:productionschema}b. The system is evacuated while the carbonyl compounds are retained in a cooling trap. 

After excess gas removal, the cooling trap is allowed to warm up to release the volatile compounds, which are then fed into the ion source and ionized in collisions with electrons. In the following sections, the feasibility of the concept is investigated by means of proof-of-principle experiments of individual steps of the full procedure and numerical simulations. Molybdenum and tungsten were used as model case for the studies. 

Despite their potential as volatile carriers, transition metal carbonyl complexes are delicate compounds, and decomposition in beam-induced plasmas and at elevated temperatures is expected \cite{Usoltsev2017Part1,Usoltsev2017Part2,Duellmann2009_presep,Wang_2014}. In addition, impurities (\textit{e.g.} oxygen or humidity) in the reactive gas reduce the chemical yield \cite{Wittwer2021a}.

The topics addressed in the following sections aim at an order-of-magnitude estimation of radioactive ion beam yields that could be achieved. 
The average expected radioactive ion beam yield $N$ computes to
\begin{equation}
\begin{split}
&N =  N_0^{\mathrm{batch}} \times \epsilon \times \nu_{\mathrm{batch}}, \mathrm{where} \\
&N_0^{\mathrm{batch}} =  \frac{I_p \; N_0}{\lambda}\left[1-\mathrm{e}^{- \lambda \: t_{\mathrm{irr}}}\right], \mathrm{and} \\
&\epsilon = \epsilon_{\mathrm{extr}} \times \epsilon_{\mathrm{stop}} \times \epsilon_{\mathrm{form}} \times \epsilon_{\mathrm{sep}}(\lambda) \times \epsilon_{\mathrm{ion}}\mathrm{,\;with} 
\end{split}
\label{eq:yield}
\end{equation}
the in-target production yield by nuclear reactions $N_0$, the fraction of isotopes propagating through foils $\epsilon_\mathrm{extr}$, and of those stopped in the gas $\epsilon_{\mathrm{stop}}$, the formation of the volatile carrier molecule $\epsilon_{\mathrm{form}}$, the conditions required for removal of excess carbon monoxide gas and the associated efficiency $\epsilon_{\mathrm{sep}}$ as well as the ionization efficiency of carbonyl complexes $\epsilon_{\mathrm{ion}}$. The number of isotopes $N_0^{\mathrm{batch}}$ with radioactive decay constant $\lambda$ is produced per batch within the irradiation time $t_\mathrm{irr}$ at steady proton current $I_p$. The repetition frequency of the batch operation is $\nu_\mathrm{batch}$. Decay during separation is included in the efficiency factor $\epsilon_\mathrm{sep}(\lambda)$. In the following sections, each parameter is explained and its numerical value is studied in detail. The considerations are based on the target concept shown in fig.~\ref{fig:productionschema} and target geometries listed in table \ref{tab:foilgeometries}. 

\subsection{In-target production}
\label{subsec:intarget}

To avoid direct exposure to the intense proton beam, a spallation neutron source made of tungsten is proposed. The latter is commonly found in ISOLDE target units and called proton-to-neutron converter \cite{Catherall2003,Stora2012,Ramos20192}. This converter is concentrically surrounded by uranium or platinum target foils, which are placed inside a carbon monoxide-containing vessel. High production rates for molybendum and other 4d transition metals are expected from uranium fission. Platinum was chosen because high production rates of tungsten are expected in spallation reactions.

 Relative to the \SI{20}{\centi\metre}-long foils, the converter is indented by 2.5 cm in the beam axis and direction. This layout is initially proposed to reduce the proton fluence, which emerges by scattering in the target container \cite{dos_Santos_Augusto_2014,Luis2012,Gottberg2014}. The geometry parameters are listed in table~\ref{tab:foilgeometries}, and chosen under consideration of typical ISOLDE target size limitations and recoil ranges, but without further optimization, which is out of the scope of this conceptual work. The geometry could be further optimized (\textit{e.g.} towards higher production rates and lower energy deposition on the tungsten rod) depending on the outcome of experiments studying carbonyl decomposition in such a setup. The power deposted by the primary proton beam might require modification of the geometry or the development of a dedicated cooling concept to avoid heating of the gas-volume to temperatures above the decomposition threshold for transition metal carbonyl compleses. The foil thickness was chosen based on expected recoil range and commercial availability. 

As will be discussed in sect. \ref{subsec:stopping}, the recoil energy of tungsten isotopes obtained in spallation reactions is lower than that of uranium fission products. Thus, a dense arrangement of thin (\SI{2}{\micro\meter}) platinum foils is desired. While the design is mechanically challenging, dense foil arrangements have already been realized by either using dimpled foils or dedicated spacers between the layers \cite{Bennett1997a}. The proposed quantity of platinum foils (\SI{26}{\square\meter}) might be cost-prohibitive or require a custom manufacturing process. Further investigations towards a different type of platinum-containing material, \textit{e.g.} films deposited on backing foils \cite{Stoychev2001}, could be subject of further studies. The replacement of platinum foils by gold foils could equally be considered.

The number of desired isotopes produced inside the target material per \si{\micro\coulomb} of primary beam $N_0$ was investigated with the FLUKA particle tracking code \cite{FLUKA1,FLUKA2}. 
The 1.4~GeV proton beam was assumed to have a Gaussian profile with $\mathrm{\sigma = 0.35\;cm}$, which is the common irradiation mode for an ISOLDE proton beam focused on the target container. 
The resulting proton and neutron fluences are shown in fig.~\ref{fig:FLUKAfluence}. 
Due to scattering of the high energy proton beam on the converter, the beam broadens to form a plume. 
Using the proton-to-neutron converter reduces the proton fluence in the region of the gas-filled container by two orders of magnitude, in comparison to direct proton irradiation. The neutrons emerge isotropically from the tungsten rod. 
Overall, the uranium foils are exposed to a neutron fluence exceeding \SI{2e11}{\per\square\centi\metre\per\second}, if the neutron converter is bombarded with \SI{2.0}{\micro\ampere} of protons.
fig.~\ref{fig:FLUKAprod} shows the in-target production yield of molybdenum and rhodium as 4d elements, and tungsten as 5d element. 
The plot contains simulation results for two different target materials and geometries, which are summarized in table~\ref{tab:foilgeometries}. 

\begin{table*}[b]
	\begin{center}
    \caption{Assumed target foil geometries for in-target production and extraction efficiencies. A sketch is shown as part of fig.~\ref{fig:productionschema}.}
    \label{tab:foilgeometries}
    \begin{tabular}{l|c|c|c} 
		& Unit & Uranium foils & Platinum foils\\
      	\hline
      	converter radius & \si{{\centi\metre}} & \multicolumn{2}{c}{\num{1.4}} \\
     	converter length & \si{{\centi\metre}} & \multicolumn{2}{c}{\num{20.0}}  \\
     	target container radius & \si{\centi\metre}& \multicolumn{2}{c}{\num{7.0}}  \\
     	\hline
	  	number of foils & & 3 & 483 \\
	  	thickness & \si{\micro\metre} & \num{25} & \num{2.5} \\
	  	length &  \si{\centi\metre} & \num{20}& \num{20} \\
	  	radii & \si{\centi\metre} & 1.9, 3.0, \num{4.0} & 1.90 to \num{6.73} \\
	  	total surface area & \si{\metre\squared} & \num{0.11} & \num{26.1} \\
    \end{tabular}
  \end{center}
\end{table*}

\begin{figure}
\resizebox{1\linewidth}{!}{%
  \includegraphics{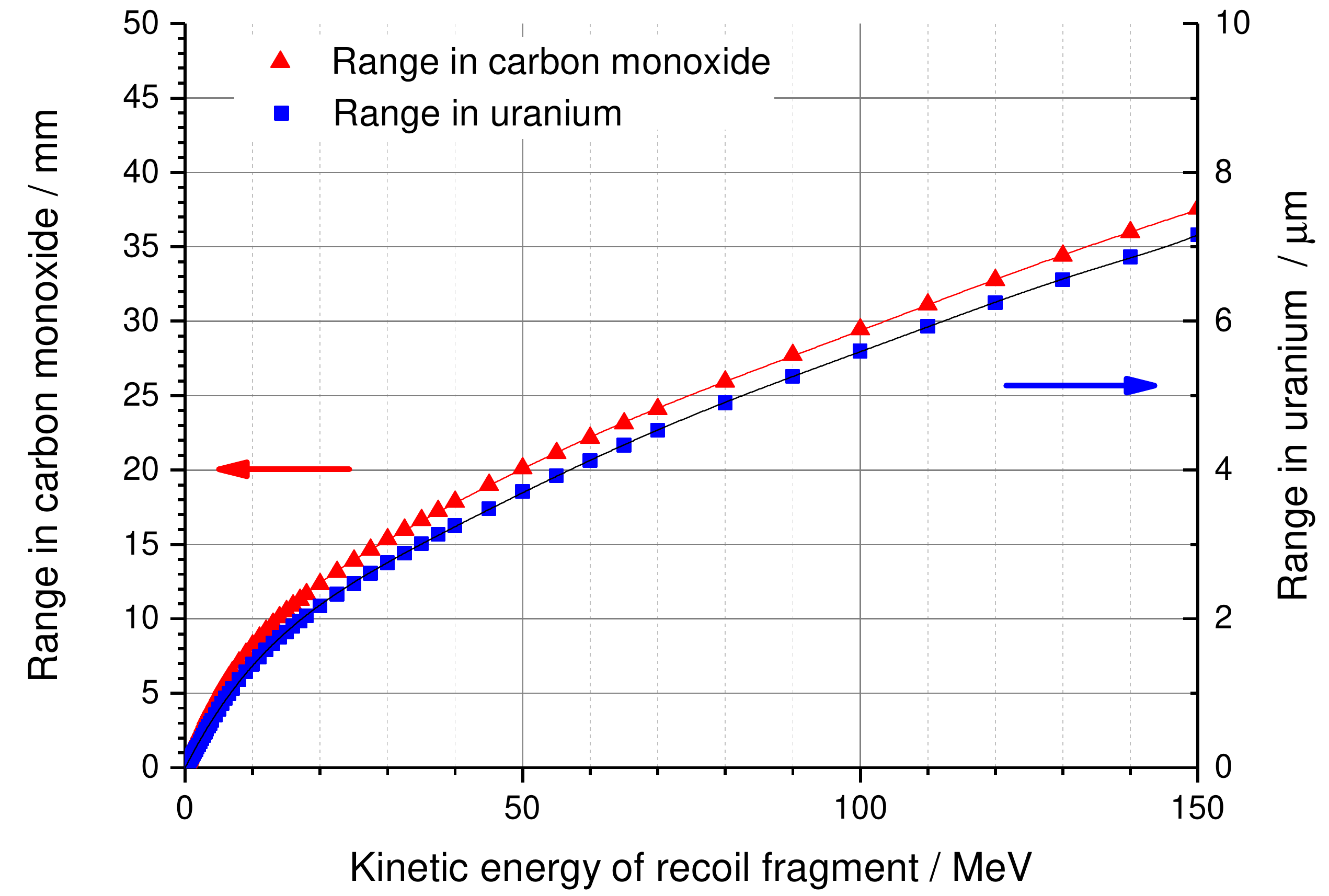}
}
\caption{Range of a \textsuperscript{105}Mo fission fragment in a carbon monoxide atmosphere at \SI{1}{bar}~(abs.) and  a metallic uranium, obtained by SRIM \cite{Srimpaper,srimonline}. The lines show a polynomial fit }
\label{fig:range}
\end{figure}

The in-target yields for molybdenum in the U(n,f) reaction reach a maximum for \textsuperscript{105}Mo ($T_\mathrm{1/2} = \SI{36}{\second}$) and decrease towards heavier and lighter isotopes. The yields on the outermost of the three foils are somewhat reduced due to the lower solid angle coverage and particle fluences. The yields of \textsuperscript{105}Mo compute to \SI{1.4e8}{\per\micro\coulomb} for the innermost foil, \SI{1.3E+08}{\per\micro\coulomb} (middle foil) and \SI{1.1E+08}{\per\micro\coulomb} (outermost foil), reaching a total yield of \SI{3.8e+08}{\per\micro\coulomb}. For comparison, the in-target production of \textsuperscript{105}Mo in a typical uranium carbide target at ISOLDE (ca. \SI{50}{\gram\per\square\centi\metre} of \textsuperscript{238}U), irradiated directly with the proton beam, is computed to \SI{5.4e+09}{\per\micro\coulomb} by FLUKA \cite{YieldDB}.

The yields of the heavier element rhodium are in the same order of magnitude as molybdenum yields.
The maximum yield of the tungsten isotopic chain in the Pt(n,spall) reaction, is found near \textsuperscript{174}W. The total yield of this isotope produced in the platinum foil assembly is predicted to be \SI{4.5E+08}{\per\micro\coulomb}.

\subsection{Recoil range, thermalization and molecule formation}
\label{subsec:stopping}


After production, the radionuclide propagates through the target material foil. Subsequently, it is crucial to thermalize the hot reaction products in the surrounding gas atmosphere to avoid implantation into the next foil or the container. The projected ranges in the respective materials depend on the kinetic energy of the recoil ion, which in turn depends on the underlying nuclear reaction, and the associated kinetic energy of the projectile. In addition, the projected range decreases with carbon monoxide pressure in the target container. In the following a pressure of \SI{1}{\bar}~(abs.) inside the target container, and a pure carbon monoxide atmosphere are assumed.

Protons and neutrons have been considered as projectiles inducing nuclear reactions, and their 
fluences in the target foils were obtained with FLUKA. Other secondary reaction channels are typically negligible and are not considered in in-target production simulations \cite{Lukic2006}. Folding the computed energy-differential projectile fluences with cross sections for isotope production computed by different codes, allows to identify the predominant reaction channel and finally, its associated recoil energy. For the low energy part (below \SI{200}{\mega\electronvolt}) of the incident particle spectra, TALYS \cite{talyspaper,talysonline} and GEF \cite{GEFreport,GEFonline} were used. Higher energies were addressed with the ABRABLA code \cite{Kelic:2009yg}. The energy distribution of spallation neutrons originating from a neutron converter in similar geometry has already been experimentally investigated and is in agreement with FLUKA simulations \cite{Stora2012}. The cross sections given by ABRABLA have also been benchmarked with experimental results obtained at ISOLDE \cite{Lukic2006,Cocolios2008}.
  
\begin{figure*} [t]
    \centering
\resizebox{1\linewidth}{!}{%
   \subfigure{\includegraphics[width=0.49\textwidth]{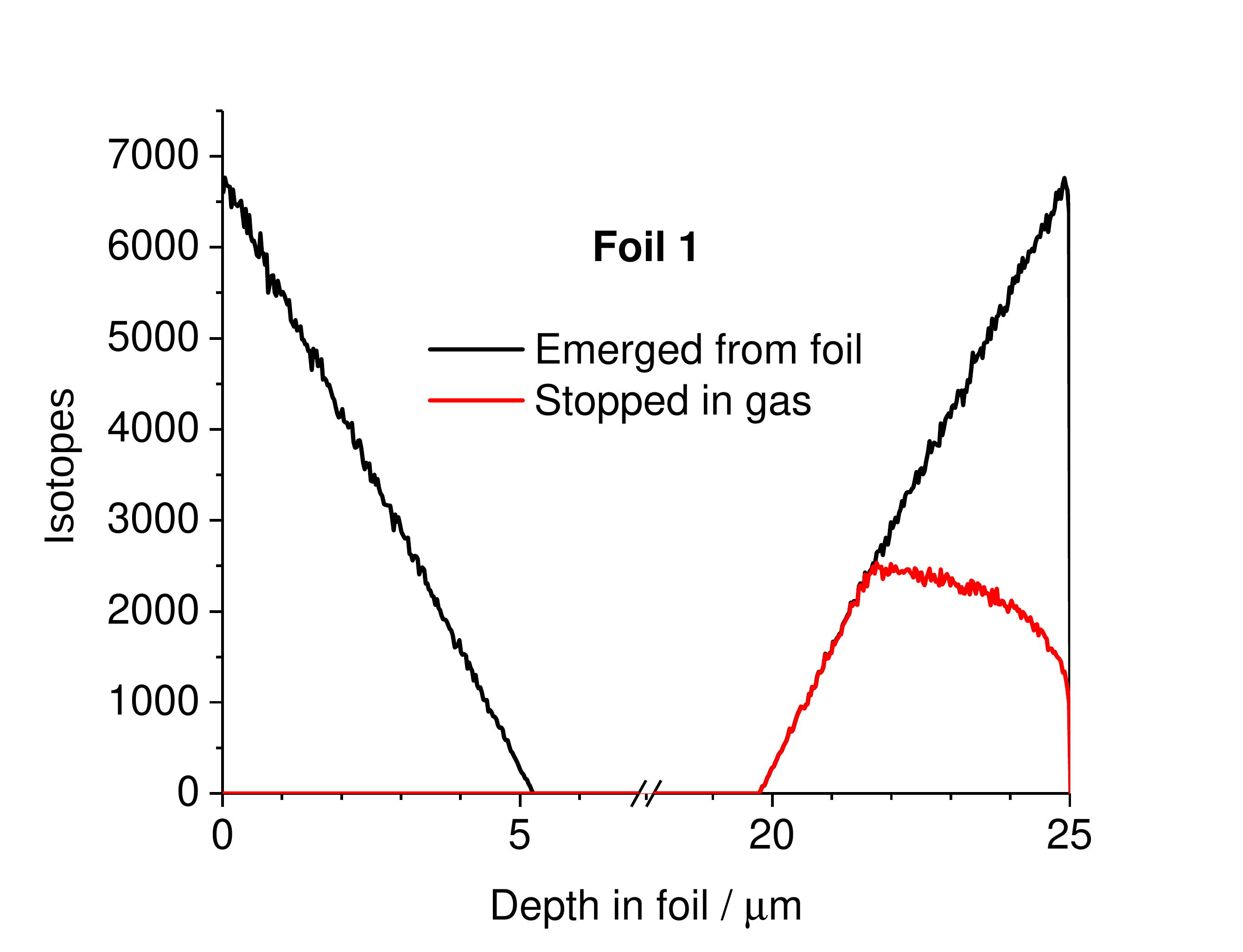}} 
    \subfigure{\includegraphics[width=0.49\textwidth]{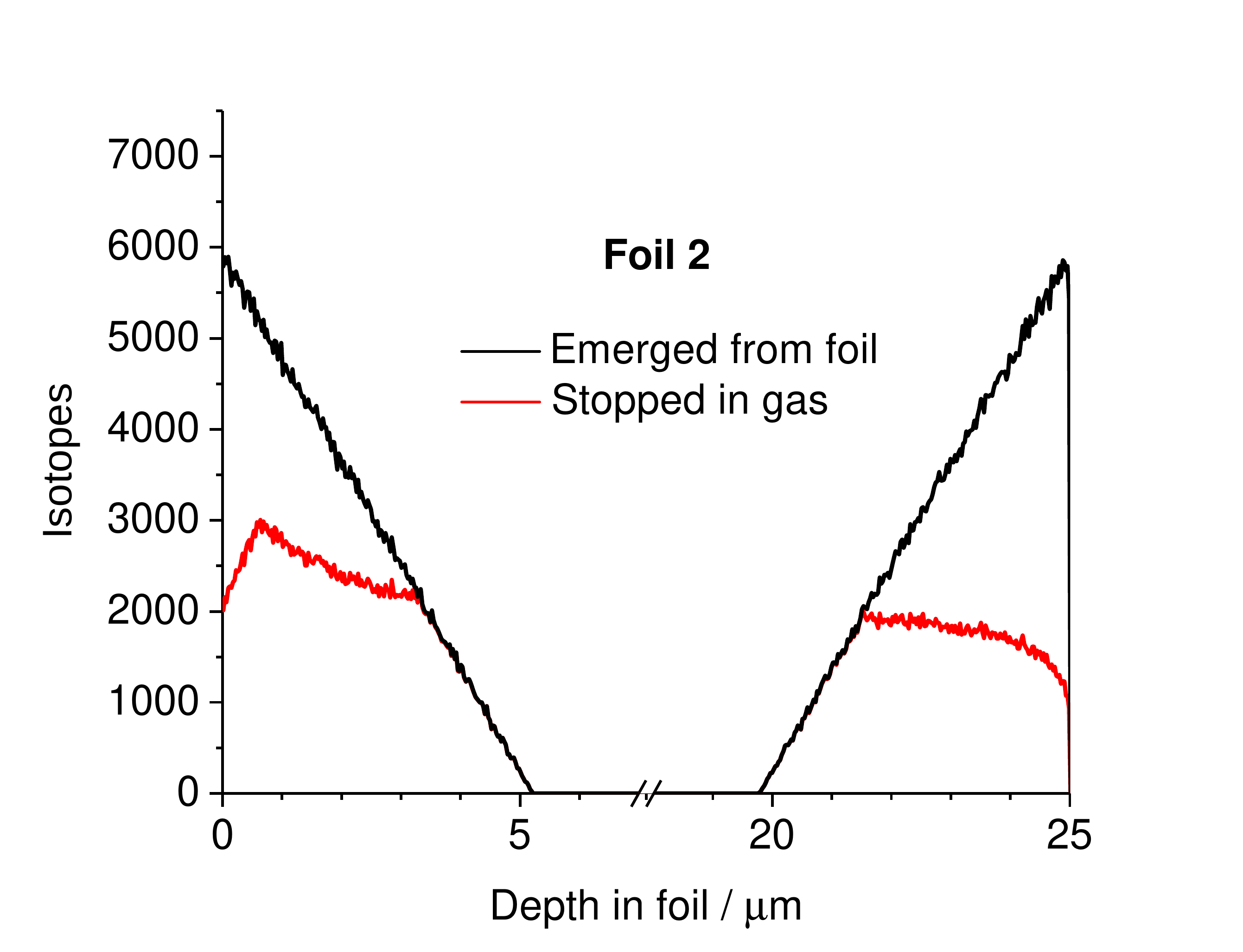}}    
        \subfigure{\includegraphics[width=0.49\textwidth]{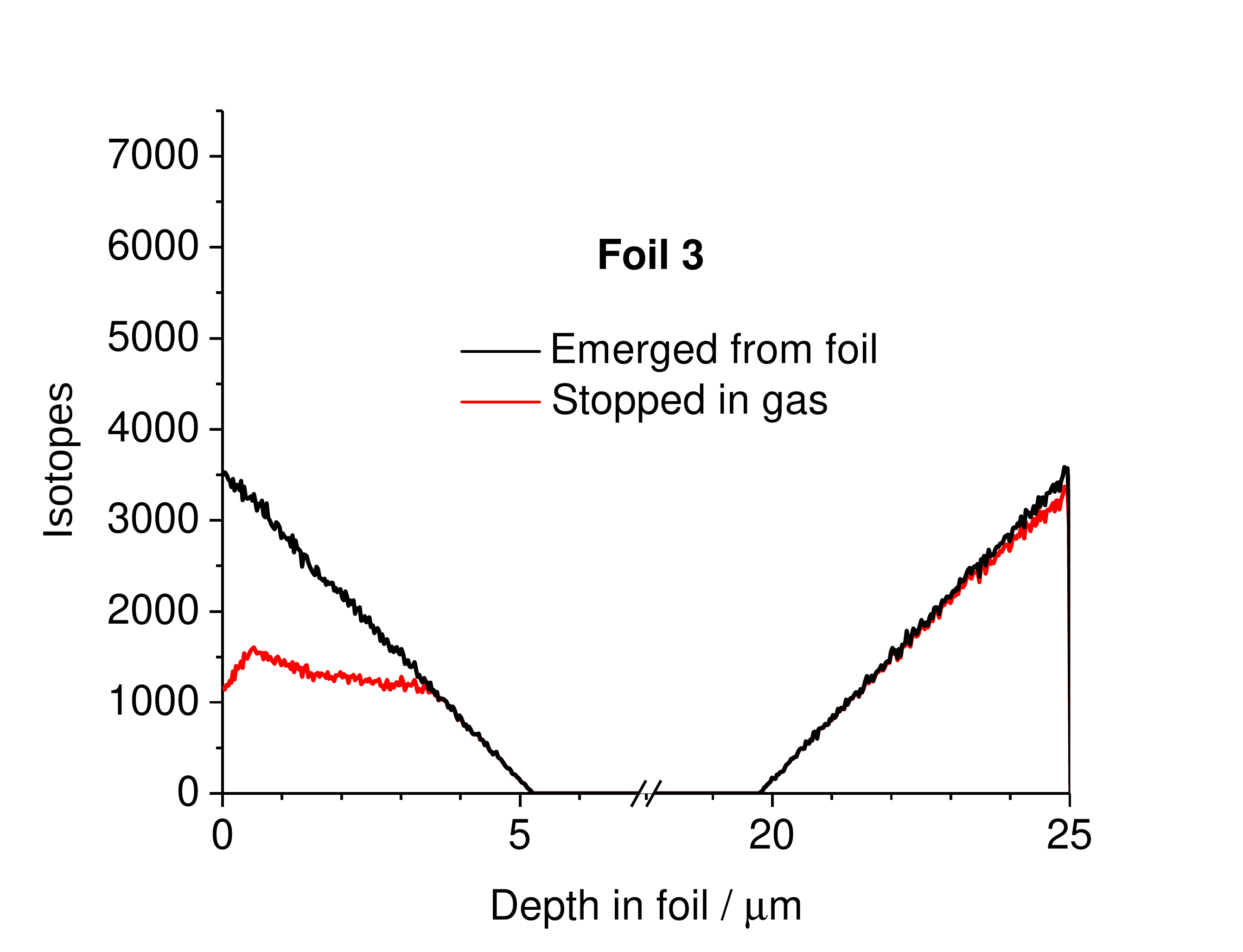}} 
    }
    \caption{Fraction of \textsuperscript{105}Mo fission recoils which emerge from a \SI{25}{\micro\metre} uranium foil, and which are stopped in a carbon monoxide gas before a confining wall is reached in dependency of the position of the fission event in the foil. A kinetic energy of 90 MeV is assumed, along with the target geometry outlined in sect.~\ref{sec:concept}. The ratio of counts in the respective foils represents the ratio of the in-target production rates obtained by FLUKA.}
 \label{fig:StoppingMC}
\end{figure*}

The neutron energy spectra obtained within this work follow a broad distribution with a maximum at ca. 3 MeV (evaporation neutron peak) and extend to ca. \SI{1.4}{\giga\electronvolt}, which is the energy of the incident proton \cite{DetlefFilges2009}. The spectra of protons  have a maximum at ca. \SI{100}{\mega\electronvolt}, and also extend to \SI{1.4}{\giga\electronvolt}.

From simulations with the GEF code in the energy range from \SI{100}{\kilo\electronvolt} to \SI{20}{\mega\electronvolt} for incident neutrons impinging on a \textsuperscript{238}U target, the mean kinetic energy of \textsuperscript{105}Mo fission recoils is expected to be between 90 and 100 MeV. ABRABLA equally predicts a recoil energy of ca. 90 MeV for 100 MeV neutrons. Excitation functions for the production of \textsuperscript{174}W from \textsuperscript{195}Pt have been calculated and folded with the proton and neutron fluence spectra. Incident particles from ca. \SI{100}{\mega\electronvolt} on, significantly contribute to tungsten production. In total, the contribution of protons and neutrons to the production of \textsuperscript{174}W, computes to 59\% and 41\%, respectively. The distribution maximum for the recoil energy of \textsuperscript{174}W is near 1 MeV. It exhibits an exponentially decreasing tail towards higher energies.

The recoil ranges in the target foils and carbon monoxide gas have been calculated with SRIM \cite{Srimpaper,srimonline} and fitted with a polynomial function. The results for molybdenum in uranium foils are shown in fig.~\ref{fig:range} and indicate that a \textsuperscript{105}Mo fission fragment has a range of ca. $\mathrm{6\; \upmu m}$ in a metallic uranium foil. 
The range of a fission fragment emerging from the uranium foil in carbon monoxide depends on its energy after it emerges from the foil. 
To account for energy losses in the uranium foil and the target geometry, Monte-Carlo simulations were performed. 
Energy losses in the foil have been estimated by inversion of the obtained range function.

Within each foil, the fission events were generated uniformly. The relative number of production events per each foil was chosen based on the data obtained by FLUKA.
The distribution of polar angle $\phi$ and azimuthal angle $\theta$ of the fission fragment trajectories were chosen such that the distribution in the solid angle $\mathrm{d}w = \mathrm{sin} \theta \; \mathrm{d} \theta \;\mathrm{d} \phi$ was uniform.\footnote{Deviations from the uniform distribution could be seen in angle distributions obtained by ABRABLA, especially in the case of tungsten produced from platinum foils. Nonetheless, the effect of non-uniformly distributed nuclear reaction recoils only decreases the extraction and stopping efficiency by no more than 10\% and is neglected. } In vector representation, the trajectory of the fission fragment is given by

\begin{equation}
\begin{pmatrix}
x \\ y \\ z
\end{pmatrix}
=
\begin{pmatrix}
x_0 \\ y_0 \\ z_0
\end{pmatrix}
+\beta
\begin{pmatrix}
\mathrm{sin} \theta \; \mathrm{cos} \phi \\ \mathrm{sin} \phi \; \mathrm{cos} \phi \\ \mathrm{cos} \theta
\end{pmatrix},
\label{eq:trajectory}
\end{equation}

where $x_0$, $y_0$ and $z_0$ are the coordinates of the fission event, and $x$, $y$ and $z$ the coordinates of intersection with a foil, after propagating a distance of $\beta$. Each uranium foil is described by two cylinders, defining the the inner and outer surface of the foil, respectively. The cylinders are defined by radius $r$ and length $\mu_\mathrm{max}$.

\begin{equation}
\forall \: \mu \in \{0,\mu_\mathrm{max}\} :
\begin{pmatrix}
x \\ y \\ z
\end{pmatrix}
=
\begin{pmatrix}
x \\ \pm \sqrt{r^2 - x^2} \\ 0
\end{pmatrix}
+\mu
\begin{pmatrix}
0 \\ 0 \\ 1
\end{pmatrix}
\label{eq:foil}
\end{equation}

Thus, equalizing the equations~\ref{eq:trajectory} and \ref{eq:foil} yields an expression for $\beta$, which represents the maximum free flight path.

The value was calculated to be 
\begin{align}
\begin{split}
\beta = &-x_0\: \mathrm{cos}\:\phi\;\mathrm{csc}\:\theta - y_0\; \mathrm{sin}\:\phi\;\mathrm{csc}\:\theta \\
&\pm \left[\mathrm{csc^2}\:\theta\: \left(r^2 - y_0^2\:\mathrm{cos}^2\:\phi \right.\right.\\
 & \left.\left. +2\:x_0\:y_0\: \mathrm{cos}\:\phi\:\mathrm{sin}\:\phi\; + \left( r^2 - x_0^2 \right) \mathrm{sin}^2\:\phi \right)\right]^{1/2}, 
\end{split}
\label{eq:travelpath}
\end{align}
under the condition that $|y_0\: \mathrm{cos}\:\phi - x_0\:\mathrm{sin}\:\phi | < r$ \cite{Mathematica}. The fraction of \textsuperscript{105}Mo fission recoils emerging from the uranium foil in dependence of the depth of the fission event location from inside the foil is shown in fig.~\ref{fig:StoppingMC}. Due to shallow angles,  only a small fraction can emerge from the foil, if the point of fission exceeds a distance perpendicular to the surface of $\mathrm{5\;\upmu m}$. Assuming a typical thickness of commercially available uranium foils of $\mathrm{25\;\upmu m}$, in total $\epsilon_{\mathrm{extr}}^{\mathrm{Mo}} = 10\%$ of all\textsuperscript{105}Mo fission recoils emerge from the foil. For comparison, a released fraction of 22\% from an uranium oxide target at \SI{1140}{\celsius} within 60 minutes was reported for Mo \cite{Eichler1975}. Release data at higher temperatures, as they are standard for uranium carbide targets, are not available and also difficult to predict due to the susceptibility of uranium oxide targets to undergo sintering.

The remaining recoil energy after propagating through the foil and the free flight path till the next surface determine if the recoil is thermalized in the gas or lost, due to implantation in a solid. The simulation predicts that $\epsilon_{\mathrm{stop}}^{\mathrm{Mo}} = 49\%$ of the emerging fragments are thermalized in carbon monoxide gas at \SI{1}{\bar}~(abs.), leading to about 5\% of all produced \textsuperscript{105}Mo fragments being thermalized in gas. The bulk target material could be reduced by integration of thinner foils of \textit{e.g.}~\SI{10}{\micro\meter} thickness, which would not decrease the total yield. Within the geometry assumed for platinum foils, $\epsilon_{\mathrm{extr}}^{\mathrm{W}} = 1.6\%$ emerge from the foil and thereof $\epsilon_{\mathrm{stop}}^{\mathrm{W}} = 21\%$ are thermalized in the gas.  

After thermalization of the recoils in carbon monoxide gas, the carbonyl compounds form readily. Measurements have shown that in gas mixtures with inert gases, the chemical yield increases with the partial pressure of carbon monoxide. Using pure carbon monoxide gas, the chemical efficiency $\epsilon_{\mathrm{form}}$ was found to be ca. 80\% for the formation of Mo(CO)\textsubscript{6} and  30\% for  W(CO)\textsubscript{6} \cite{even2014situ}.\footnote{The given efficiency for molybdenum was obtained as ratio between transport by carbonyl formation and aerosol transport of atomic species attached to clusters. An absolute efficiency is given for tungsten.}

%

\subsection{Gas separation}
\label{subsec:gassep}

A carbon monoxide partial pressure of 1 bar (abs.) inside the target container is desirable to achieve a high chemical yield, and is also crucial for efficient stopping of the fission recoils. However, ion sources typically operate under high vacuum conditions, and often a pressure of \SI{1e-3}{\milli\bar}   may already prevent efficient ionization. Thus, the separation of radioactive carbonyl compounds from the excess gas atmosphere is required. Since carbon monoxide, exhibiting a boiling point of \SI{-191}{\degreeCelsius} \cite{CRChandbook}, is by far more volatile than carbonyl compounds, the two components are separable by chromatography. The implementation of a suitable cryogenic trap coupled to an ion source was already evaluated for the transport of carbon monoxide in a carrier gas by Powell \textit{et al.} \cite{Powell_1998}, and Katagiri \textit{et al.} have also shown the feasibility of the integration into the process of radioactive ion beam production \cite{Katagiri2015}. The option of neutral CO injection via a cryogenic trap into an electron beam ion source for charge breeding is also considered for medical applications \cite{Pitters2020,Boytsov2015,pitters2018summary}.

Adsorption enthalpies of carbonyl compounds have been measured on SiO\textsubscript{2}, gold, Fluorinated Ethylene Propylene (FEP) and PolyTetraFluoroEthylene (PTFE) surfaces by isothermal chromatography and by thermochromatography \cite{DissJulia,Wang_2014,Cao2016,Wang2015}. In isothermal studies with radiotracers from a \textsuperscript{249}Cf source, an adsorption enthalpy for Mo(CO)\textsubscript{6} on SiO\textsubscript{2} surfaces of $-\Delta H_{\mathrm{ads}} = \SI{42.5 \pm 2.5}{\kilo\joule\per\mole}$ was determined  \cite{even2014situ}. In thermochromatography experiments, the same  quantity was found to be $-\Delta H_{\mathrm{ads}} = \SI{36 \pm 8}{\kilo\joule\per\mole}$. Later experiments by Wang \textit{et al.} reported values in agreement with earlier findings. Adsorption enthalpies of other transition metal carbonyl complexes (technetium, tungsten, rhenium,  osmium, iridium) could also be deduced, all ranging in the same order of magnitude and indicating a physisorption interaction. For the following considerations, Mo(CO)\textsubscript{6} was chosen as a model case, and an adsorption enthalpy of \SI{-40}{\kilo\joule\per\mole} was assumed as average value of the measurements. 


The system for cryogenic gas separation must be designed such that the total gas flow rate into the ion source does not exceed its maximum acceptable gas load and the pumping capacity of the vacuum system. 
A typical upper limit of \SI{1e-3}{\milli\bar\liter\per\second} was estimated for gas injection into an ion sources operated at ISOLDE. In addition, the time needed to reach this condition should be as low as possible to minimize decay losses. The proposed setup is shown in fig.~\ref{fig:productionschema}b, which is operated in a batch-mode. During irradiation, the target container is filled with carbon monoxide at atmospheric pressure. After a defined irradiation time, which depends on the half-life of the desired isotope, the gas inventory is pumped by a roughing pump through a cooled quartz tube which retains the less volatile carbonyl compounds. After a certain fraction is evacuated from the target container, the latter is isolated from the chromatographic system by closing the reservoir valve. The residual pressure in the chromatographic tube is further reduced, till the maximum flow rate into the ion source is reached. Finally, the quartz tube is heated by a resistive heating element to ambient temperature allowing carbonyl compounds to be fed into the ion source, where they are ionized and electrostatically extracted. The time required for allowing the temperature to raise, is assumed to be short in comparison to the half-life of the model isotopes discussed within this concept ($\ll \SI{35}{\second}$) and neglected in the calculation of the expected yield. In parallel to the heating of the quartz tube, the irradiation of the next batch can take place, so that the heating period is not included in the cycle time $\nu_\mathrm{batch}^{-1}$ of the batch process.

\begin{figure}
\resizebox{1\linewidth}{!}{%
  \includegraphics{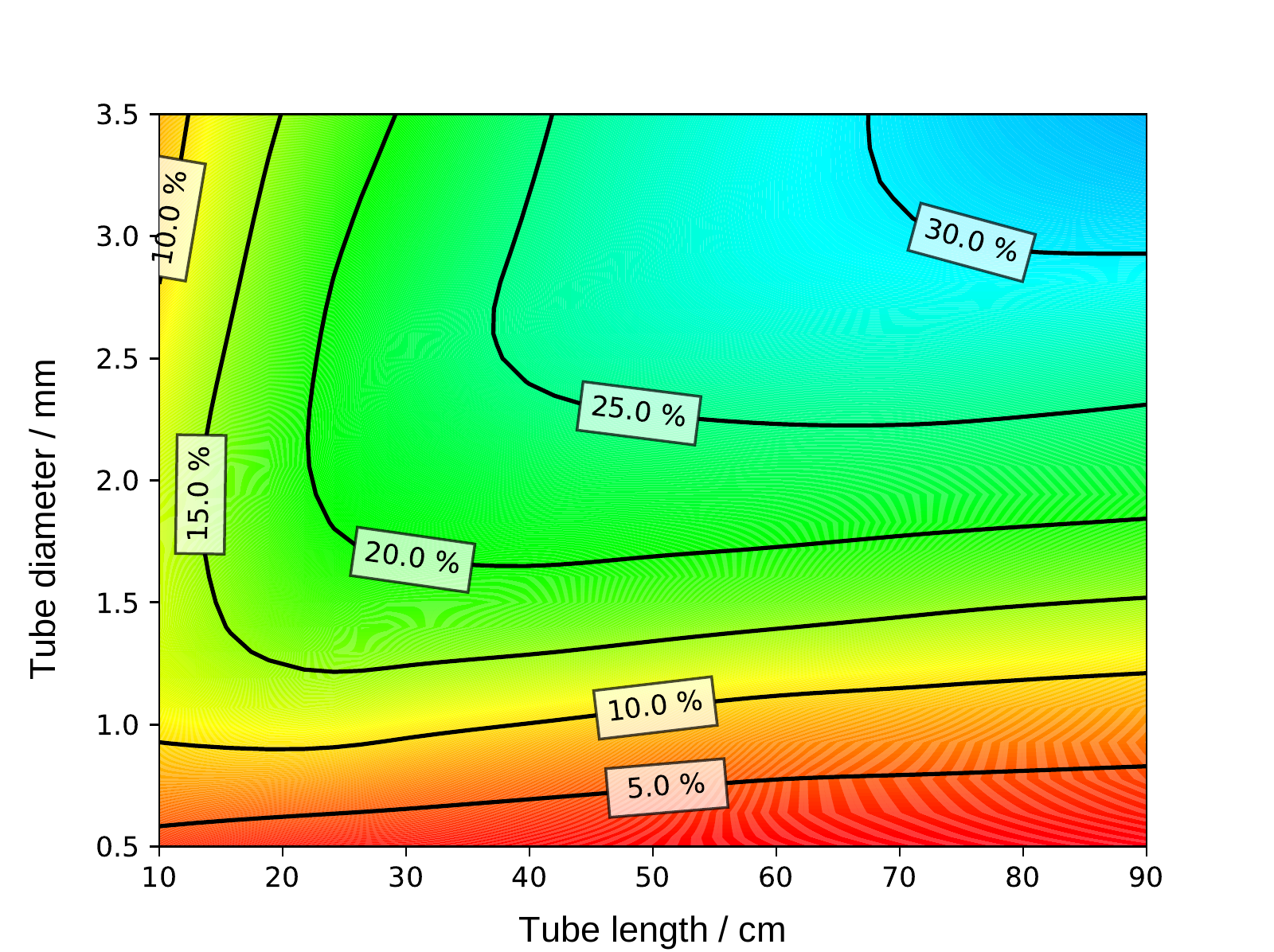}
}
\caption{Dependence of the cryogenic gas separation efficiency $\epsilon_\mathrm{sep}$  on the dimensions of the chromatographic tube for \textsuperscript{105}Mo, which is determined by decay losses during extraction of the reservoir inventory and consideration of irreversible processes obtained by simulation (see text). The figure shows interpolated data.}
\label{fig:pumpeff}
\end{figure}

A Monte-Carlo model proposed by Zvara \cite{ZvaraOrig} was used to investigate the feasibility of the concept, which was also similar to the model used by Even \textit{et al.} for the adsorption enthalpy measurements \cite{DissJulia}. The model takes into account the temperature dependent sojourn times (cf. eq.~\ref{eq:frenkel}) and gas flow conditions. Assuming that the chromatographic tube is connected to an evacuated vessel (choked flow), the evaluation of Reynolds numbers for tube diameters from \SI{0.5}{\milli\metre} to \SI{3.5}{\milli\metre} suggests turbulent flow conditions. The time dependent pressure in the reservoir $p_{\mathrm{target}}(t)$ can be estimated following the evaluation of the pV-flowrate $q_{pV}$, given by 

\begin{equation}
q_{pV} = V \frac{\;\mathrm{d}p}{\mathrm{d}t} = A \; a \; p_{\mathrm{crit}}\;,
\end{equation}

where $V$ is the volume of the target container, $A$ the cross sectional area of the tube, $a$ the speed of sound in the medium and $p_{\mathrm{crit}}$ the critical pressure, which is given by

\begin{equation}
 p_{\mathrm{crit}} = 1.92 \; \frac{1}{a\;d} \left( \frac{\overline{c}^{\:6}}{\eta}\right)^{1/7} \left(\frac{d^3\;p_{\mathrm{target}}^2}{2\;l}\right)^{4/7}\; ,
\end{equation}

where $d$ and $l$ are inner diameter and length of the tube, respectively, $\overline{c}$ the mean thermal particle speed and $\eta$ the temperature-dependent dynamic viscosity, which was calculated with the Jones equation \cite{johnston1942viscosities}. It was assumed that the tube was kept at a constant temperature. The equations hold under the approximation that $p^2_{\mathrm{target}} \approx p^2_{\mathrm{target}} - p^2_{\mathrm{crit}}$ \cite{Wutz}. Solution of the equations yields the time-dependent pressure in the reservoir, which was initially ($t=0$) kept at the pressure $p_{\mathrm{target,0}} $.

\begin{equation}
p_{\mathrm{target}}(t)=\frac{823543 \; p_{\mathrm{target, 0}} \; V^7 }{\left( k \; p_{\mathrm{target, 0}}^{1/7} \; t + 7V\right)^7 } 
\end{equation}

The simulation generates particles with a random lifetime, which is sampled from a distribution according to the given half-life. Subsequently, the propagation of the particle through the chromatographic system is simulated. The length of a displacement is typically approximated by the variance of the zone profile of the chromatographic peak, and expressions have been obtained for laminar flow \textit{e.g.}~ref.~\cite{giddings1965}. Within this work, the mean length of a displacement $\delta$ was approximated by the diffusional deposition length in developed turbulent flow, which depends on Reynolds number $Re = Q\: d \: \rho / ( A \: \eta)$, Schmidt number $Sc = \eta / ( \rho \: D)$, the mutual diffusion constant $D$ and the volume flowrate $Q$. The density $\rho$ is derived from the pressure, and the diffusion constant is obtained by the approximation proposed by Gilliland \cite{Gilliland1934}. Assuming a uniform distribution of particles at the inlet, the expression for the mean displacement length holds for the first contact with a wall and was sampled from an exponential distribution \cite{Zvra2008}.

\begin{equation}
\delta = \frac{43.5\;Q}{\pi \: D \: Re^{0.83}\; Sc^{0.3}}
\label{eq:jumplength}
\end{equation}
 
 According to Zvara, the mean number of real wall collisions within a long displacement step $Z$ can be calculated by the flowrate, the mean thermal particle speed and the  surface per unit length $S$ of the column as

\begin{equation}
Z = \frac{\overline{c}\;S}{4\;Q} \; \delta.
\end{equation}

After each displacement, the time spent by the particle in the system and its position are evaluated. At the end of the time required to extract a defined fraction from the reservoir, it is computed, if the particle has either decayed, was retained on the column surface, or has been eluted. 

The separation efficiency is given by a linear combination of i) the fraction of carbonyl compounds evacuated from the reservoir ii) the ratio of molecules which are inside the quartz column (\textit{i.e.} not decayed and not eluted) after the desired fraction was extracted from the reservoir, to the number of molecules fed into the column and iii) an additional factor to account for decay losses during the time required to reach a flow rate below $10^{-3}$~\si{\milli\bar\liter\per\second}, which was estimated to be ca. 5 seconds. The input variables of the simulation were the fraction of carbonyl compounds extracted from the reservoir and fed into the quartz tube as well as its dimensions and temperature.

For each geometry of the separation channel the respective highest efficiencies $\epsilon_\mathrm{sep}$ have been determined. At elevated flowrates, longer channels are required due to the increase in the diffusional deposition length (eq.~\ref{eq:jumplength}). The results for \textsuperscript{105}Mo are shown in fig.~\ref{fig:pumpeff}. The simulation predicts that efficiencies above 60\% are in reach.  Using the same geometry boundary conditions, the maximum efficiency for \textsuperscript{108}Mo (1.11~s) is calculated to be above 0.6\%. Typical temperatures are in the range of $-130 \pm 40 $ \degree C. Further details of the Monte-Carlo simulation are discussed in ref.~\cite{Zvra2008}. 

In addition to radioactive decay, further losses have to be considered, such as irreversible sticking after decomposition of the volatile compound or reactions with impurities. The extent of such additional losses can be estimated based on typical capillary transport losses, which could be in the order of 50\% \cite{Even_2012}. Decay losses for the long-lived isotope \textsuperscript{174}W ($T_{1/2} = \SI{31}{\minute}$) are negligible, and an efficiency of $\epsilon_{\mathrm{sep}}^{\mathrm{W}} = 50\% $ is assumed. The separation efficiency of \textsuperscript{105}Mo computes to $\epsilon_{\mathrm{sep}}^{\mathrm{Mo}} = 30\% $.

\subsection{Ionization}
\label{subsec:ion}

A review about ion sources for radioactive ion beam production can be found in ref.~\cite{Stora:1693046}. In contrast to ion sources designed to deliver stable isotope beams, additional requirements arise for the ionization of radionuclides. Due to their limited availability compared to stable isotopes, the ionization efficiency is one of the most important figures of merit. For exotic isotopes with very short half-lives ($\lesssim \SI{100}{\milli\second}$), the residence time also needs to be considered. A compact design is required to meet constraints imposed by robot-handling of the target and ion source unit and resistance to the strong radiation field of the driver beam is mandatory.

The three main processes for ion generation are i) surface ionization ii) photo-induced ionization and iii) electron-impact ionization. Surface ionization in hot cavities is applied for elements with low ionization potential (IP) of up to ca. \SI{6}{\electronvolt} like the alkaline or alkaline earth metals. For elements with elevated IP, resonant laser ionization can be used if laser systems are available to excite suitable transitions \cite{Marsh:1967371}. Electron impact ionization is the underlying process in electron beam (arc-discharge) ion sources and radio-frequency driven plasma ion sources. Via electron impact ionization, almost all elements and molecules can be ionized efficiently. Plasma sources are typically used for volatile species only.


The ionization of carbonyl compounds is a crucial step towards ion beam production. The method of ionization must be chosen carefully with respect to the properties of the compound. The ionization potential of Mo(CO)\textsubscript{6} was measured to be in the range from 8.2 eV to 8.5 eV \cite{Chen1997,Masuda1992,Bursten1984,Michels1980}. The first bond dissociation energy (FBDE) is significantly lower and was determined to be 1.7~eV by pyrolysis with a pulsed CO\textsubscript{2}-laser in a gas cell \cite{lewis1984organometallic}, in agreement with data obtained by thermal decomposition on a silver 
surface \cite{Usoltsev2017Part1,Usoltsev2017Part2} and theoretical studies \cite{ehlers1994structures}. In comparison to typical candidates for molecular beams at ISOLDE, the compound is delicate, and decomposition on hot surfaces is expected. For species with high ionization potential, Forced Electron Beam Induced Arc-Discharge  (FEBIAD) ion sources \cite{kirchner1976investigation}, like the VADIS (Versatile Arc-Discharge Ion Source) \cite{Penescu2010} are commonly used. However, the high operating temperature is expected to decompose the carbonyl compounds even before they reach the ion source volume. Thus, ion sources operated below decomposition temperature, favoring high electron energies for ionization and efficient ion extraction over breakup, are the preferred choice.

\subsubsection{ECR sources}
The Metal Ions from VOlatile Compounds (MIVOC) meth\-od, where volatile metallic compounds are fed into an Electron Cyclotron Resonance (ECR)-heated plasma ion source, is a well established method for the extraction of non-ra\-dioactive metal beams \cite{Brown:1113300,Koivisto1994}. The residence time of an element in the ion source is an important parameter for radioactive ion beam sources. This data is not yet available for MIVOC ion sources.

The production of molybdenum beams by injection of molybdenum hexacarbonyl is reported by Nakagawa \textit{et al.} \cite{nakagawa1997production} at the RIKEN 18 GHz ECRIS. The reported ion currents $I_\nu$ on each charge state $\nu$ along with the material consumption allow a rough estimation of the ionization efficiency. For ferrocene, the material consumption $q_\mathrm{m}^\mathrm{Fe} = \SI{2}{\milli\gram\per\hour}$ is explicitly reported. Based on the respective vapour pressures of molybdenum hexacarbonyl $p_\mathrm{Mo}$  and ferrocene $p_\mathrm{Fe}$, the material consumption of Mo(CO)\textsubscript{6} can be estimated here, under the assumption that the control valve between MIVOC chamber and ion source was adjusted to the same conductance. The ionization efficiency is given by 


\begin{equation}
\epsilon_\mathrm{ion} = \frac{\sum_{\nu} I_\nu / \nu }{e\:q_{\mathrm{n}}} = \frac{M_\mathrm{Fe} \sum_{\nu} I_\nu / \nu }{q_\mathrm{m}^\mathrm{Fe} \; \mathrm{F} } \frac{p_\mathrm{Fe}}{p_\mathrm{Mo}}\:,
\end{equation}

where $e$ is the elementary charge, $q_{\mathrm{n}}$ the neutral particle flow into the ion source, $M_{\mathrm{Fe}} = \SI{55.8}{\gram\per\mole} $ the molar mass of iron and F the Faraday constant. Assuming vapour pressures at \SI{300}{\kelvin} of \SI{1}{\pascal} for ferrocene \cite{Silva1990}, and \SI{25}{\pascal} for molybdenum hexacarbonyl \cite{baev1980thermodynamic}, ion currents, and ferrocene material consumption as given in ref.~\cite{nakagawa1997production}, the ionization efficiency of Mo(CO)\textsubscript{6} computes to \SI{0.03}{\percent}. The ionization efficiency for iron from ferrocene computes to 4.3\%. Ionization efficiencies could possibly be improved by using a gas mixture as buffer gas \cite{Geller:381888}. Given that currents were not reported for all charge states, the calculated efficiency is a lower limit. 


\subsubsection{Arc-discharge ion sources}

In arc-discharge ion sources, electrons are emitted from a cathode and accelerated to an energy of ca. \SIrange{100}{200}{\electronvolt} that matches the maximum ionization cross section. The electrons are typically emitted thermionically. Arc-discharge ion sources, in contrast to compact radio-frequency plasma ion sources, offer the advantage of a narrow electron energy distribution and relatively high electron energies. The latter avoid favouring breakup over ionization due to insufficient electron energy. Thus, the application of electron impact ionization in a cold environment presents an interesting asset. Penescu \textit{et al.} proposed eq.~\ref{eq:VADISbasiceff} to model the ionization efficiency of FEBIAD-type ion sources  \cite{LiviuDiss}. The ionization efficiency $\epsilon_\mathrm{ion}$ depends on the rate of ionization per unit volume $R_\mathrm{ioniz}$, the volume of the ionization region $V$, the number of neutral particles injected per unit time $n_\mathrm{in}$ and an additional factor $f$ to account for the probability of ion extraction, electron confinement and higher order effects. The rate of ionization can be expressed as a linear combination of the number densities of neutral particles $N_n$, electrons $N_e$, the ionization cross section $\sigma$ and the relative velocity of electrons and neutral particles $v_\mathrm{el}$.

\begin{equation}
  \epsilon_\mathrm{ion} = \frac{R_\mathrm{ioniz} V }{n_\mathrm{in}}  f \mathrm{,  \;\;where \;\;} R_\mathrm{ioniz} = N_n \: N_e \: \sigma \: v_{\mathrm{el}}
  \label{eq:VADISbasiceff}
\end{equation}

Unfortunately, absolute partial ionization cross sections for molybdenum hexacarbonyl or its fragments have not yet been published. It is interesting to note that also the ionization cross section of molybdenum has not yet been measured due to the refractory nature of the element and only theoretical calculations are available \cite{Kwon2005}. 
\begin{figure}
\resizebox{1\linewidth}{!}{%
  \includegraphics{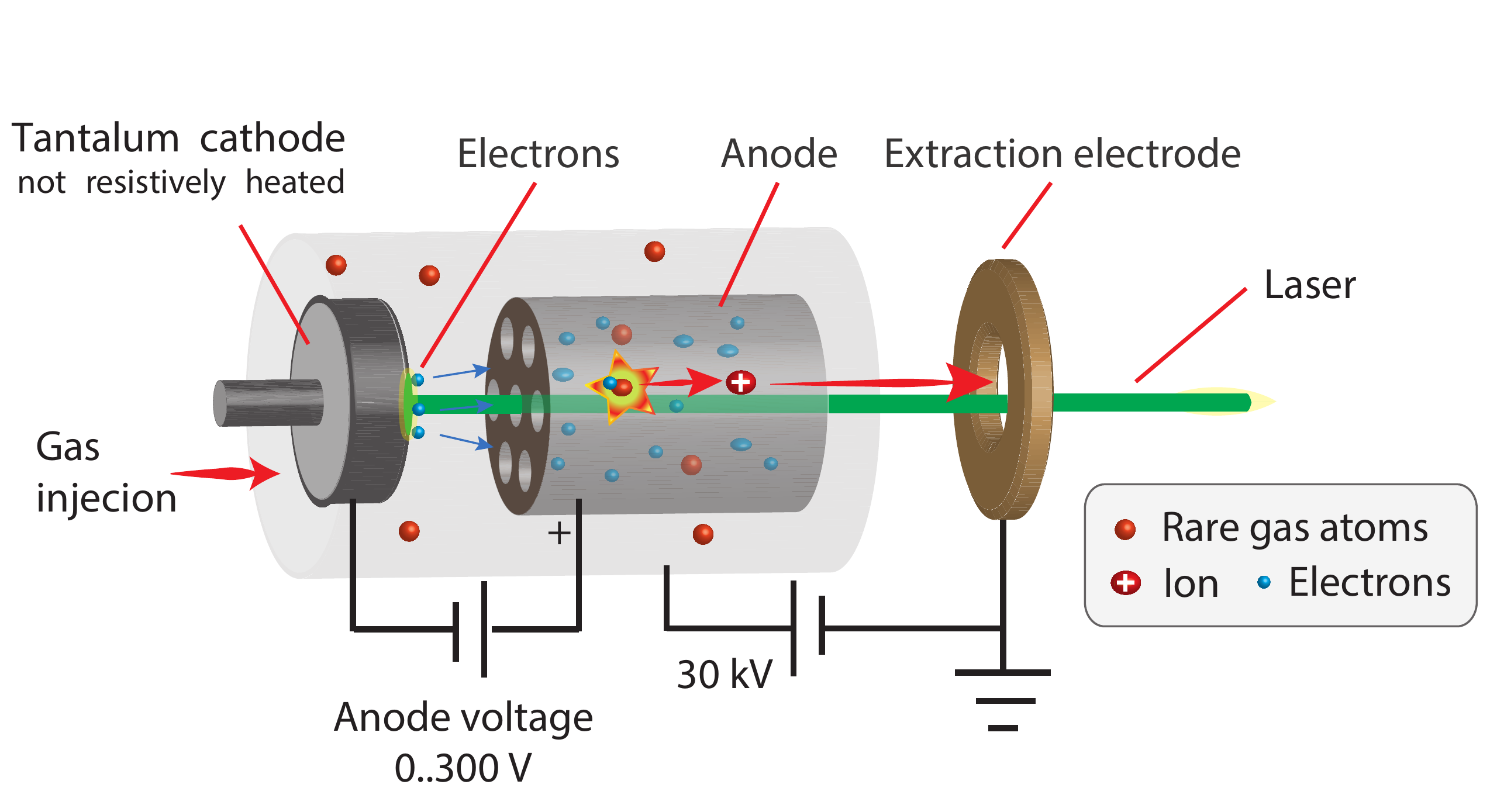}
}
\caption{Principle of the exploratory experiment exploiting electron generation by a laser beam in a cold environment. The electrons are liberated from the cathode and accelerated through a grid into the positively biased anode body, where volatile species are ionized in collisions. }
\label{fig:photocathode}
\end{figure}

Due to the lack of available data, we conducted a comparative study of the ionization efficiency of the noble gas krypton and molybdenum hexacarbonyl in a cold environment. We chose to use a laser to liberate electrons out of the tantalum cathode of a standard VADIS source equipped with a water cooled transfer line (VD7). In contrast to common operation of the VADIS ion source, the source was not resistively heated. The laser pulses impinging on the cathode induce a local raise of temperature during the laser pulse. However, the heat is dissipated quickly to the water-cooled assembly \cite{Jaeger1953}. The laser power used in our experiment is not expected to raise the average temperature of the cathode significantly. Thus, thermal decomposition of carbonyl compounds on the cathode is negligible.

The candidate mechanisms for electron generation in the interaction of the laser beam and the tantalum cathode are either extraction from a plasma plume, thermionic electron emission or the photo-electric effect. As discussed in the next paragraph, the used fluences were most likely insufficient for ablation and plasma formation. Time-de\-pen\-dent heat transfer and thermionic electron emission models for tantalum are available \cite{Verber1965}, but require more precise knowledge of pulse fluence as available from our experiment to be applied. Since the single photon energies were below the material work function, the absorption of multiple photons is required to release electrons by the photo-electric effect. Efficient multiphoton photoemission has been observed with ultrashort ($\tau_{\mathrm{p}} = \SI{80}{\femto\second}$) laser pulses \cite{Musumeci2010} but data on quantum efficiency is not available for the conditions present in our experiments.  Nonetheless, in both scenarios electrons are extracted in an environment that is (on average) at ambient temperature. 

\paragraph{Experimental}

A sketch of the ion source is shown in fig.~\ref{fig:photocathode}. UV light (\SI{343}{\nano\metre}) supplied by a Pharos laser at \SI{50}{\kilo\hertz} repetition rate and pulse length of \SI{265}{\femto\second} was guided through the ion beam outlet aperture on the tantalum cathode. The laser power before entering the vacuum system was measured to be \SI{4.5}{\watt}. The dimensions of the laser spot were estimated with the bare eye and measured to be ca. \SI{5}{\milli\meter} in diameter. Due to the limited ion beam outlet aperture of only \SI{1.5}{\milli\meter} in diameter, only a fraction of the beam power reached the cathode. An upper limit for the fluence per pulse computes to $\phi_l < \SI{5e-3}{\joule\per\square\centi\meter}$. The minimum laser fluence needed for ablation (threshold fluence) was estimated in ref.~\cite{Mittelmann2020} to be \SI{0.17}{\joule\per\square\centi\meter} at $\lambda_l = \SI{750}{\nano\meter}$ and a pulse length of $\tau_{\mathrm{p}} = \SI{8.5}{\femto\second}$. Thus, the source in our experiment was most likely not operated in an ablation regime.\footnote{Attempts to further focus the laser beam with a telescope caused high voltage breakdowns and damage on the cathode surface. However, in the experiments described in this work, the focusing telescope was not used.}


Krypton (Carbagas, 99.998\%) and molybdenum hexacarbonyl (Schuchardt M\"unchen, TA Mo 36.33\%, C 27.18\%, Fe 0.005\%, Cu 0.0008\%) were supplied through a common transfer line into the ion source. The krypton flow rate was controlled with a calibrated leak, which was measured to be \SI{1.15e-5}{\milli\bar\liter\per\second} for \SI{1}{\bar} (abs.) of helium. The setup for the controlled injection of molybdenum hexacarbonyl consisted of an evaporation chamber, connected to the common transfer line via a regulation valve (Pfeiffer EVR116). The evaporation chamber was equipped with a capacitance diaphragm gauge (Pfeiffer CMR 373) to monitor the pressure. The residual gas composition was monitored by a residual gas analyzer (Pfeiffer PrismaPlus) at the extraction site of the ion source.

\begin{figure}[b]
\resizebox{1\linewidth}{!}{%
  \includegraphics{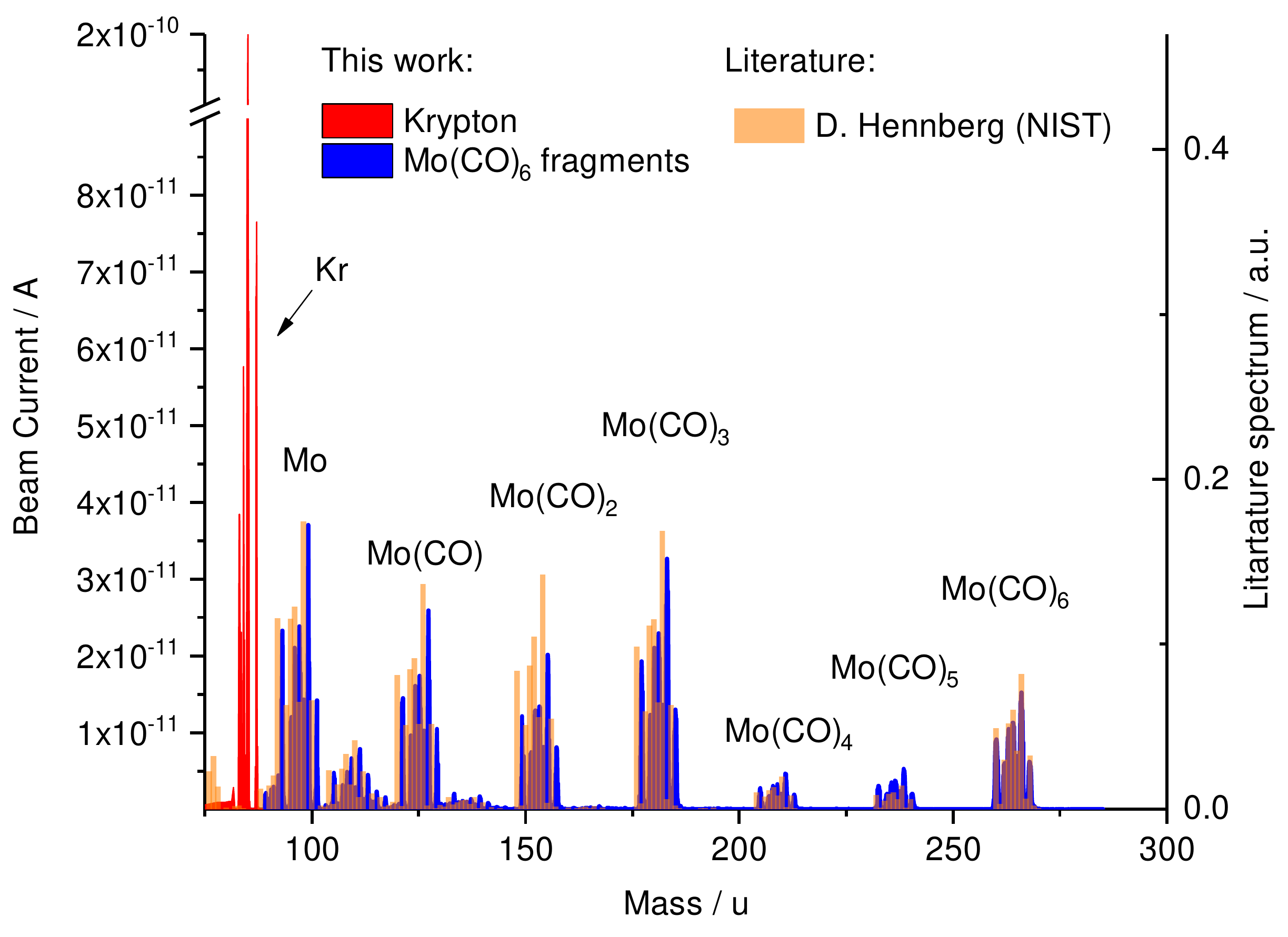}
}
\caption{Typical mass spectrum of Kr and Mo(CO)\textsubscript{6} measured simultaneously with the cold electron impact source. A mass spectrum of Mo(CO)\textsubscript{6} obtained also by electron impact ionization by D. Henning in \cite{NISTWebbook} is shown for comparison. Minor deviations in the fragment distribution might arise from different electron energies or source characteristics.}
\label{fig:massspec}
\end{figure}

\begin{table}
	\begin{center}
    \caption{Ionization efficiency estimation of the exploratory cold electron impact ion source shown in fig.~\ref{fig:photocathode}. The efficiency is given for the most abundent fragment of Mo(CO)\textsubscript{6}, and corrected for isotope abundance. The aim of the experiment was to deduce the \textit{relative} ionization efficiency of the two species.}
    \label{tab:photocathodeeff}
    \begin{tabular}{c|c|c|c|c}
    	\multirow{2}{*}{Nuclide} & Isotope &  \multirow{2}{*}{Current} & Injected & \multirow{2}{*}{Ionization} \\
	   	    	& abundance &   & neutrals &  efficiency \\
    	\hline
    	\rule{0pt}{3ex}  
    	
    	\textsuperscript{98}Mo 		& 24.3\%	&	\SI{33}{\pico\ampere} & \SI{2238}{\particle\nano\ampere} & \SI{0.0015}{\percent} \\
    	\textsuperscript{84}Kr 		& 57.0\%	&	\SI{200}{\pico\ampere} & \SI{8999}{\particle\nano\ampere} & \SI{0.0022}{\percent} \\

    \end{tabular}
  \end{center}
\end{table}


\begin{table*}
	\begin{center}
    \caption{Parameters of a proposed cold electron-impact ion source that exploits a photo-cathode as source of electrons and is expected to reach an ionization efficiency of 1\% for Mo(CO)\textsubscript{6}. The ionization efficiency of the proposed source is estimated by scaling the measured ionization efficiency of Mo(CO)\textsubscript{6} obtained in an exploratory experiment by the respective calculated electron currents. In the experiment, electrons were liberated by a laser, however the exact mechanism of electron production could not be identified. The increase in electron current (\textit{cf.} eq.~\ref{eq:MaxCurrent}) can be achieved by adaptation of the geometry which allows higher bunch charges without virtual cathode formation and an increase in the repetition rate of the laser. An anode potential of $V_g = \SI{120}{\volt}$ is assumed. See text for details.}
    \label{tab:photocathodeeff}
    \begin{threeparttable}
	    \begin{tabular}{l|c|c|c|c}
	    	Parameter & Symbol & Unit & Exploratory Experiment & Proposed Photo-cathode source \\
			\hline
	    	\multicolumn{5}{l}{}\\
	    	\multicolumn{5}{l}{1. Geometry and calculated critical bunch charge} \\
	    	\hline
	    	laser spot diameter \tnote{\textdagger} & $2\:r_\mathrm{em}$ & \si{\milli\meter} & \num{1.5} & \num{12} \\
	    	anode-cathode distance & $d_{\mathrm{g}}$ & \si{\milli\meter} & 1.5 & 3 \\
	    				critical bunch charge \tnote{\S}& $q_{\mathrm{c}}$  & \si{\pico\coulomb} & 2 &  45 \\
			\multicolumn{5}{l}{}\\
	    	\multicolumn{5}{l}{2. Cathode and laser wavelength} \\
	
	    	\hline
	    	cathode material & & & Ta & Cu \\
			center wavelength & $\lambda_{\mathrm{l}} $ & \si{\nano\meter} & 343 & 257 \\
			quantum efficiency at $\lambda_{\mathrm{l}}$ & $\epsilon_{\mathrm{q}}$ & \% &  -- \tnote{\textdaggerdbl}  & 0.014 \cite{SrinivasanRao1991} \\
	    	
	    	\multicolumn{5}{l}{}\\
	    	\multicolumn{5}{l}{3. Required laser system} \\
	    	\hline

			pulse repetition rate & $\xi$ & \si{\mega\hertz} & 0.05 & 2 \\		
			
			average power & $P_{\mathrm{l}} $ &  \si{\watt} &  $< \num{4.5}$ \tnote{$\|$} & 3.7 \\
			pulse energy & $E_{\mathrm{l}}$ & \si{\micro\joule} & $< \num{90}$ \tnote{$\|$} & \num{1.9} \tnote{$\!$\#} \\
			fluence per pulse & $ \phi_{\mathrm{l}} $ & $\si{\micro\joule\per\square\centi\meter}$ & $< \num{5093}$ \tnote{$\|$} & \num{1.6} \\
				    	
			\multicolumn{5}{l}{}\\
	    	\multicolumn{5}{l}{4. Calculated mean electron current and ionization efficiency} \\
	    	\hline
	    	mean electron current & $I_{\mathrm{m}}$ & \si{\micro\ampere} & 0.09 & 90 \\
	    	ionization efficiency & $\epsilon_{\mathrm{ion}}$ & \% & 0.001 & 1 \\
	
			\multicolumn{5}{l}{}\\

		\end{tabular}
		\begin{tablenotes}[flushleft]
			\item[\textdagger] circular area on the cathode that is illuminated and emits electrons\vspace{0.1cm} \\
			\item[\S] \parbox[t]{\dimexpr\linewidth-1em}{\linespread{0.8}\selectfont calculated maximum charge that can be transported in a short pulse between cathode and anode without virtual cathode formation} \vspace{0.1cm} \\
			\item[\textdaggerdbl] data on multiphoton quantum efficiency not available and mechanism of electron production unclear \\
			\item[$\|$] An upper limit is given based on the measured laser power before entering the vacuum system. \\
			\item[$\!$\#] Chosen to match critical charge $q_{\mathrm{c}}$
		\end{tablenotes}
    \end{threeparttable}
  \end{center}
\end{table*}

After introduction of a solid Mo(CO)\textsubscript{6} sample into the evaporation chamber, the latter was evacuated with a turbo-molecular pump, backed by a dry scroll pump, to a pressure below \SI{1e-2}{\milli\bar}. Subsequently, the valve to the pumping group was closed. The sample quickly evaporates untill the saturation vapour pressure is reached inside the reservoir. Successive opening of the regulation valve allowed controlled injection into the transfer line. The material consumption was estimated by allowing the complete evaporation of a known amount, which was monitored via the pressure of the evaporation chamber and by the residual gas composition. The material consumption was measured to be $\SI[parse-numbers=false]{88^{+\:5}_{-21}}{\micro\gram\per\hour}$. The relatively large error is due to consideration of material losses during the initial evacuation.

The results of the relative efficiency measurement of Kr and Mo(CO)\textsubscript{6} are listed in table~\ref{tab:photocathodeeff} and an obtained mass spectrum is shown in fig.~\ref{fig:massspec}. The ionization efficiencies of krypton and molybdenum compute to \SI{0.0022}{\percent} and \SI{0.0015}{\percent}, respectively. The result of the experiment shows that the ionization efficiencies of Kr and Mo(CO)\textsubscript{6} are in the same order of magnitude. While these ionization efficiencies in combination with high in-target production rates, or vaporization of a radioactive Mo(CO)\textsubscript{6} sample obtained by other means, would already allow a range of nuclear physics experiments, a higher efficiency is desirable. 

\paragraph{Proposed design of a cold photo-cathode driven ion source}

In the following, we explore the feasibility of an electron-impact ion source exploiting the photo-electric effect for electron release. Basic design parameters for a novel photo-cathode driven ion source will be derived that aims at an ionization efficiency of $\sim 1\%$ for Mo(CO)\textsubscript{6}. First, the general assumptions are presented, then the space-charge limitations of an electron current passing through the gap between cathode and anode are discussed and finally the basic requirements of a suitable laser system are given.

The efficiency estimate of the new design is based on the assumption that the measured ionization efficiency obtained in our exploratory experiment linearly scales with the electron current. This is expected in a first approximation, since the rate of ionization linearly depends on the electron density. The probability of ion extraction, which is included besides other effects in the factor $f$ of equation~\ref{eq:VADISbasiceff}, is affected by source geometry. However, in contrast to the required increase of several orders of magnitude on the electron current, the extraction factor is typically only affected to some minor extend \cite{LiviuDiss}, which could be subject of further studies.

The electron currents in our experiment have been obtained as drain current of the anode power supply. However, the instantaneous currents might have exceeded acceptable values for the used pico-amperemeter. To avoid underestimating the electron current, which would lead to overestimation of the photo-cathode source efficiency, we assume a space-charge limited current in following considerations. The measured electron currents were about factor two below the calculated space-charge limited current at a typical anode potential of \SI{120}{\volt}. The observation of saturation effects of electron current with laser pulse energy might also indicate operation in proximity of a space-charge limited regime.

The maximum current density $J_{\mathrm{p}}$ that can pass a gap of length $d_{\mathrm{g}}$ with a potential difference of $V_{\mathrm{g}}$ is classically given by the Child-Langmuir law \cite{Child1911,Langmuir1913}.  If the transition time of electrons between cathode and anode grid is long compared to the laser pulse of duration $\tau_{\mathrm{p}}$, a single sheet approximation can be applied \cite{Valfells_2002}. Corrections for the two-dimensional geometry have been proposed in ref.~\cite{Lau_2001}. The maximum mean current $I_{\mathrm{m}}$ (without virtual cathode formation occurring) is given by
\begin{equation}
  I_{\mathrm{m}} = J_{\mathrm{p}} \: \pi r_\textrm{em}^2 \: \xi \: \tau _{\mathrm{p}} = \left( 1 + \frac{d{_\mathrm{g}}}{4 r_\textrm{em}} \right) \frac{\epsilon_0 \pi r_\textrm{em}^2 V_\mathrm{g}}{d_\mathrm{g}}  \: \xi,
  \label{eq:MaxCurrent}
\end{equation}
where $\xi$ is the pulse repetition rate and $r_\textrm{em}$ the radius of the electron-emitting surface. To allow higher electron currents, the repetition rate of the laser or the diameter of the electron-emitting surface can be increased. Increasing the anode potential $V_{\mathrm{g}}$ would push the ionization cross section into an unfavorable regime, and the further reduction of anode-cathode distance (typically \SI{1.5}{\milli\meter}) reduces the reliability of the ion source since a minor displacement is sufficient to cause an electrical contact between anode and cathode.\footnote{Such issues were observed for resistively heated cathodes of VADIS at ISOLDE.} 

The laser driver is most effective if the resulting electron pulse charge does not the exceed the critical limit $q_\mathrm{c} = I_{\mathrm{m}} \xi^{-1}$. The characteristic parameters, such as mean power $P_{\mathrm{l}}$, pulse energy $E_{\mathrm{l}} = P_{\mathrm{l}} \xi^{-1}$ and pulse energy fluence $\phi_{\mathrm{l}} = E_{\mathrm{l}} \pi^{-1} r_{\mathrm{em}}^{-2}$ can be estimated by the quantum efficiency $\epsilon_{q}$, \textit{i.e.} the ratio of emitted electrons to photons hitting the surface, as

\begin{equation}
	E_{\mathrm{l}} = \frac{h\:c}{\lambda_{\mathrm{l}}} \frac{q_{\mathrm{c}}}{\epsilon_{\mathrm{q}} \: e} ,
  \label{eq:pulseenergy}
\end{equation}
where $h$ is the Planck constant, and $\lambda_{\mathrm{l}}$ the photon wavelength. Many high-quantum-efficiency materials are semiconductors, like cesium telluride. They require sophisticated vacuum systems (pressures below \SI{e-9}{\milli\bar}) to reach their nominal performance \cite{Kong1995a}. On the other hand, metal cathodes have a lower quantum efficiency but can operate at higher pressures. Thus, a copper cathode was chosen for the estimations which has a quantum efficiency of ca. \SI{0.014}{\percent} at \SI{266}{\nano\metre} and a residual pressure of \SI{e-7}{\milli\bar}  \cite{SrinivasanRao1991,Kong1995a}. 


A set of design parameters for a cold photo-cathode driven ion source and laser system is proposed in table~\ref{tab:photocathodeeff}. Following the previous discussion, the parameters are chosen such that an ionization efficiency of 1\% for Mo(CO)\textsubscript{6} could be reached under the assumption that the measured ionization efficiency scales linearly with the space-charge-limited electron current. The proposed design assumes a laser spot diameter of $2r_{\mathrm{em}} = \SI{12}  {\milli\meter}$ which is equal to the size of the VADIS cathode. In our preparatory experiment (\textit{c.f.} fig.~\ref{fig:photocathode}), the laser beam was guided through the ion beam outlet hole on the cathode. An increase of the outlet hole diameter from currently \SIrange{1.5}{12}{\milli\meter} might significantly decrease the source efficiency because it reduces the residence time of neutral species in the anode body. Thus, we propose to introduce the laser beam perpendicular to the electron beam, a trajectory that was recently developed for the perpendicularly-illuminated LIST ion source \cite{Heinke2016}. The aforementioned laser path is estimated to require an increased anode-cathode distance of ca. \SI{3}{\milli\meter}. The required pulse energy fluence computes to \SI{1.9}{\micro\joule\per\square\centi\meter} which is well below the ablation threshold of \SI{0.77}{\joule\per\square\centi\meter} even for short (\SI{10}{\femto\second}) pulses \cite{Nathala2016}. Other factors contributing to damage of photo-cathodes have been identified. In ref.~\cite{Zheng2017}, the damage threshold for copper photo cathodes was estimated by simulation and a fluence of less than \SI{40}{\milli\joule\per\square\centi\meter} is recommended. The reliability and efficiency of such a photo-cathode source might be impacted by condensation of molecule fragments on the cathode or the residual pressure of carbon monoxide. Its behavior needs to be experimentally verified.

\begin{figure*}[t]
	\centering
     \subfigure[Extraction without source magnet]{
     	\begin{overpic}[width=0.45 \textwidth]{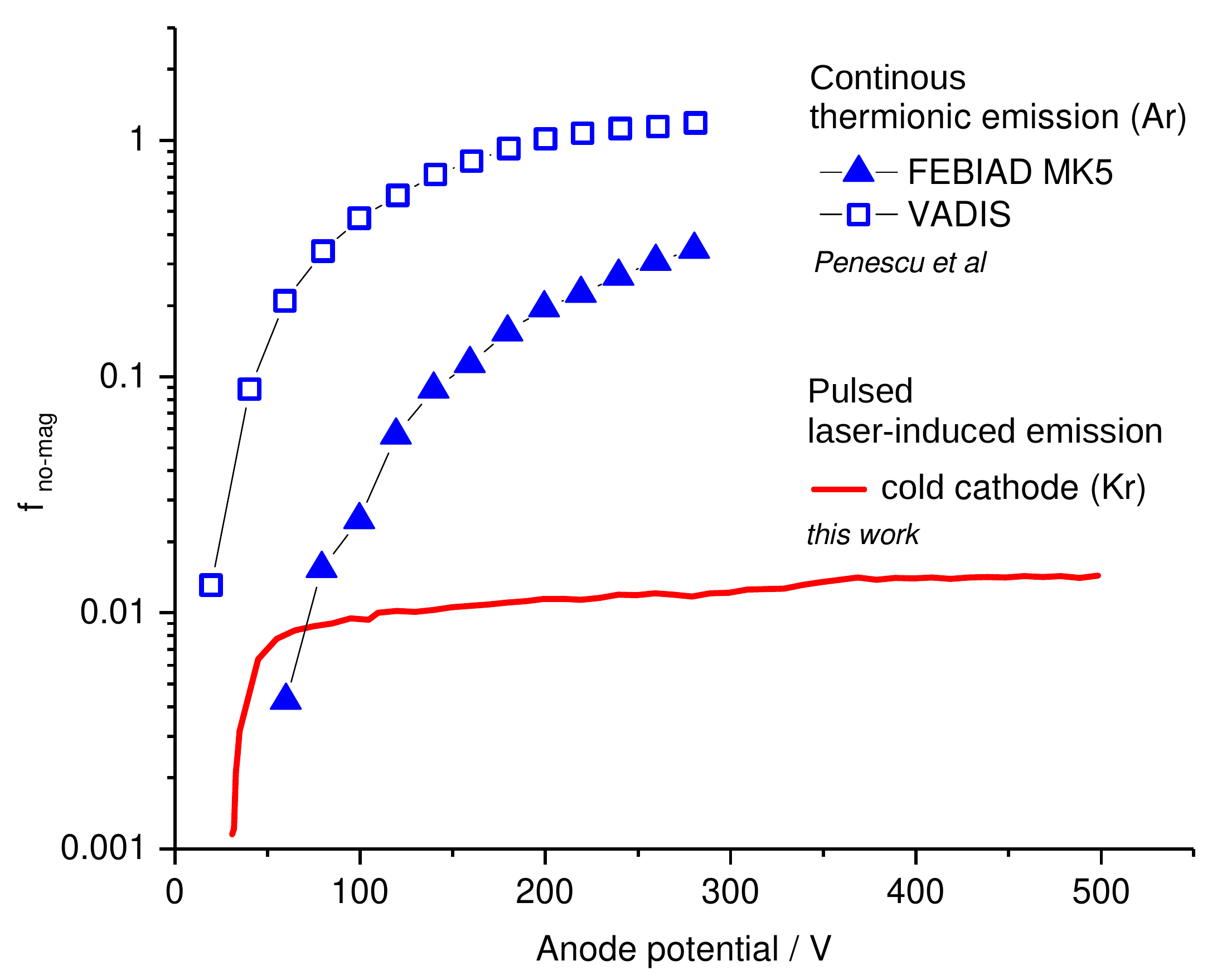}
     	\put (82,58) {\scriptsize\cite{LiviuDiss}}
     	\end{overpic}
     	}
    \hspace{0.6cm}
    \subfigure[Contribution of source magnet]{
	    \begin{overpic}[width=0.45 \textwidth]{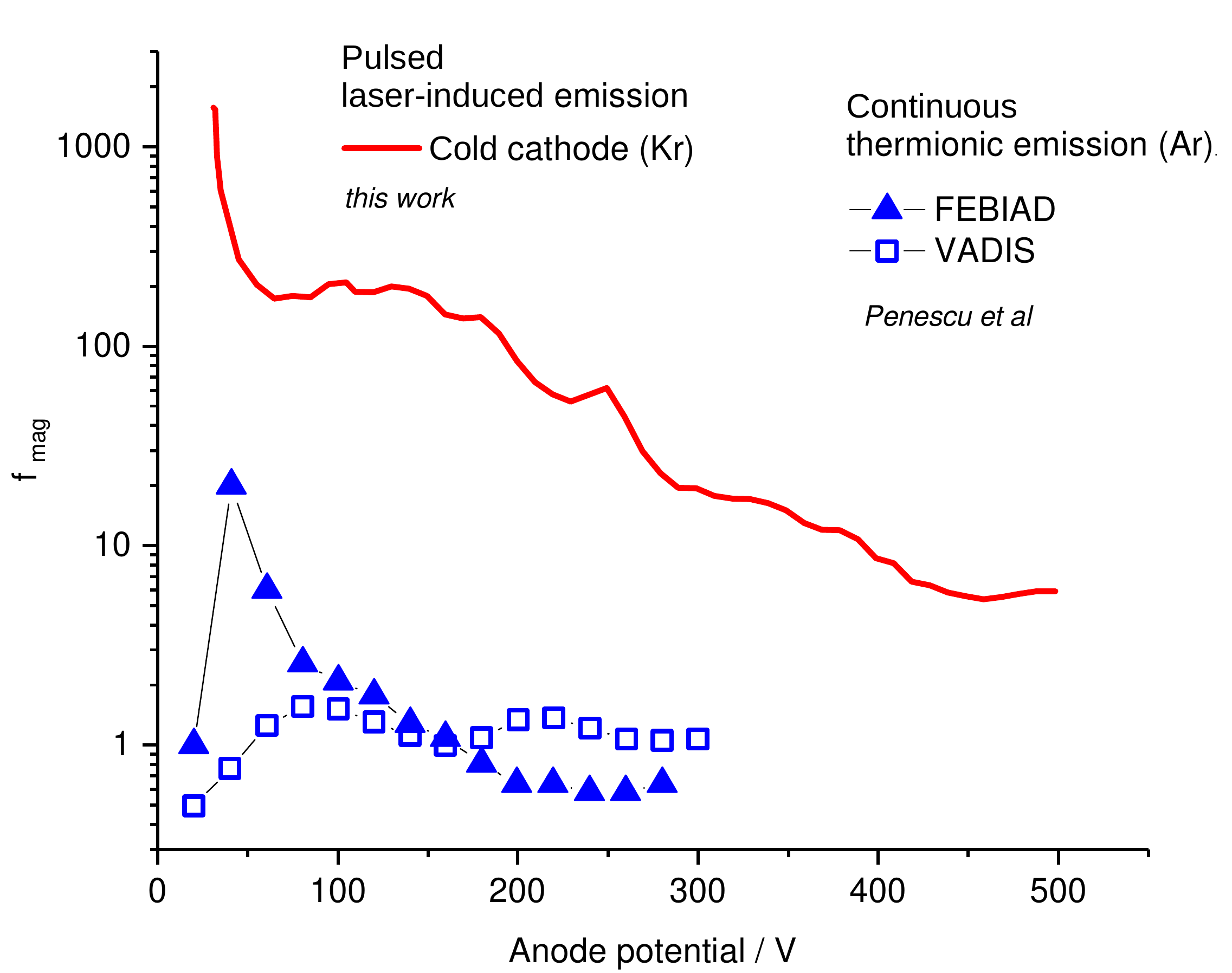}
	    \put (84.5,52.5) {\scriptsize\cite{LiviuDiss}}
     	\end{overpic}
    }
    \caption{Contributions to the extraction factor $f = f_\mathrm{no-mag}\: f_\mathrm{mag}$, according to eq.~\ref{eq:VADISEffpract}, for the exploratory cold ion source shown in fig.~\ref{fig:photocathode}   used in this work, in comparison to data from FEBIAD sources, which were taken from \cite{LiviuDiss}. The magnet currents in (b) were \SI{4.5}{\ampere}  and \SI{6}{\ampere} for the continuous thermionic source and the pulsed laser-induced emission source,  respectively. See text for details. }
 \label{fig:ffactor}
\end{figure*}

A deep ultra violet (DUV) laser is required for efficient release of electrons from metal photo-cathodes. Due to the limited bunch charge, high repetition rates  are beneficial. As listed in table~\ref{tab:photocathodeeff}, the desired efficiency of ca. 1\% for Mo(CO)\textsubscript{6} could be reached with \SI{3.7}{\watt} average power at a wavelength of \SI{257}{\nano\meter} and repetition rate of \SI{2}{\mega\hertz}. A recent review about DUV laser generation is given in ref.~\cite{Xuan2018}. The required laser system could be based on fourth harmonic generation of a \SI{1}{\micro\meter} Yb fiber laser. Besides pulsed lasers, also continuous wave (cw) lasers could be considered \cite{Zhao2017,Burkley2019,Burkleydiss}. For the latter,
additional considerations apply for ionization efficiency estimates, as will be discussed in the following paragraphs.

\paragraph{Continous thermionic source efficiency} The spatial separation of a resistively heated cathode and ionization volume provides an alternative path towards a cold electron impact ion source. A thermionic electron source could be placed remotely in an actively cooled environment, with no line of sight to the ionization volume to avoid radiative heating. The achievable ionization efficiencies of Mo(CO)\textsubscript{6} in such a configuration can be estimated from krypton efficiencies, which are in the range of 30\% for known FEBIAD-type sources \cite{kirchner1976investigation,Penescu_2010}. 

For this, our experimental results are compared to the aforementioned model (eq.~\ref{eq:VADISbasiceff}). While the ionization cross-section of a given compound is independent of the ion source, the extraction factors $f$ of a continuously operated thermionic emission ion source and a pulsed laser-induced emission source are expected to differ, even if the geometry of the cavity is similar. In the first case, the continous release of electrons generates an electric field in the ionization volume which influences the extraction of ions. It was proposed that this is due to the formation of a potential well \cite{Penescu_2010,MartinezPalenzuela:2672954,Millan2020,Palenzuela2018}. In comparison to the immediate extraction of a nascent ion guided by a  favourable field, a potential well might also increase the number of wall collisions of an ion before extraction. The latter would decrease the extraction factor $f$, particularly for condensible species which are lost to surfaces upon collision.  The pulsed electron generation in our experiments is expected to significantly reduce the aforementioned hindrance due to the limited life-time of electrons \cite{LiviuDiss}. On the other hand, the electric field present in sources with continuous electron emission might guide produced ions towards the outlet. In an electron-free environment, the operation at room temperature contributes to a more efficient ion extraction. The field produced by the grounded extraction electrode penetrates into the anode volume and decreases the potential near the outlet aperture. Ions created in this region are guided along the decreasing field towards the outlet hole. To overcome the gradient to the outlet hole, a certain ion energy is required. Thus, as argued in ref.~\cite{LiviuDiss}, the region of direct extraction (active volume) decreases with increasing ion energy.

By evaluation of the parameters given in eq.~\ref{eq:VADISbasiceff} similar to the derivation discussed in \cite{LiviuDiss}, it can be written as 

\begin{equation}
\epsilon_\mathrm{ion} = \frac{I}{\mathrm{e}\, S_\mathrm{out}} \: \sqrt{\frac{2\,\pi\,M}{\mathrm{R}\,T}}\: \sigma \:l \:f, 
\label{eq:VADISEffpract}
\end{equation}
where $I$ is the electron current, $\mathrm{e}$ the elementary charge, $S_{\mathrm{out}}$ the cross-sectional area of the outlet hole, $T$ the temperature of the ion source, $M$ the molar mass of the neutral species, $\mathrm{R}$ the universal gas constant, $l$ the length of the ionization volume in the axial direction and $\sigma$ the ionization cross-section. As in eq.~\ref{eq:VADISbasiceff}, the probability of ion extraction (and other effects) is solely included in the factor $f$. As already discussed, the electron current $I$ was approximated with the theoretical space charge-limited current, estimated for short pulses \cite{Valfells_2002} and corrected for the limited emitter surface \cite{Lau_2001}. 

The dependence of the extraction factor $f$ on the anode potential is shown in fig.~\ref{fig:ffactor} for krypton beams from the exploratory cold ion source (\textit{cf.} fig.~\ref{fig:photocathode}) using evaluated cross sections \cite{Higgins:203081}, along with data obtained for argon from FEBIAD-type sources \cite{LiviuDiss}. For this measurement, light at a wavelength of \SI{515}{\nano\metre} was used. The extraction factor is given as linear combination $f = f_{\mathrm{no-mag}}\:f_{\mathrm{mag}}$.
The factor $f_\mathrm{no-mag}$ is directly obtained from a measurement with disabled source magnet. The factor $f_{\mathrm{mag}}$ is calculated by dividing the extraction factor with enabled magnet by the previously obtained factor $f_{\mathrm{no-mag}}$. 
As can be seen from fig.~\ref{fig:ffactor}, the similarity in the curves for pulsed laser-induced and continous thermionic electron emission indicates that in both cases electron impact ionization is observed. As discussed in more detail in ref.~\cite{LiviuDiss}, the factor $f_{\mathrm{no-mag}}$ increases with anode potential, which can be attributed to a reduced space-charge  repulsion at higher electron velocities leading to a better electron confinement.
For FEBIAD-type sources, the ion extraction might also be favored at higher voltages due to a reduced Debye-length and increased active volume. The magnetic field increases the electron density,  an effect which is significantly more pronounced for the laser-induced electron emission ion source. Deviations between the cold exploratory ion source and FEBIAD-type sources arise not only due to the laser induced, pulsed release of electrons, but also due to a different geometry of electron extraction. 
 While the whole cathode surface (ca.  \SI{12}{\milli\metre} diameter) emits electrons in thermionic mode, the laser spot is defined by the diameter of the outlet aperture of the source, which is \SI{1.5}{\milli\metre} only. Seeing the similar shape of the curves, an order-of-magnitude estimation of the extraction factor for the \FEBIAD{} source in the ionization process of molybdenum hexacarbonyl is proposed in the following, provided the cathode heating and the ionization volume with the anode can be decoupled as already discussed.

Due to unknown ionization cross sections of Mo(CO)\textsubscript{6}, the extraction factor $f$ cannot be experimentally obtained from an efficiency measurement. However, the short lifetime of electrons after a laser pulse, in comparison to longer extraction times of ions, suggests that the hindrance of extraction by electrostatic fields inside the cold exploratory ion source is low. Thus, as first approximation it is assumed that the krypton extraction factor $f_\mathrm{Kr}^\mathrm{L}$ is equal to the extraction factor of molybdenum hexacarbonyl $f_\mathrm{Mo(CO)_6}^\mathrm{L}$ in the cold exploratory ion source.\footnote{A mass dependence of the extraction factor has been observed, however even the ratio of xenon and argon extraction factors was found to only be $\cfrac{f^{\mathrm{VD}}_{\mathrm{Xe}}}{f^{\mathrm{VD}}_{\mathrm{Ar}}} \approx \num{1.8}$ \cite{LiviuDiss}.} The situation in the \FEBIAD{} source is different, and the extraction factor of Mo(CO)\textsubscript{6} is approximated with the extraction factor of carbon monoxide, which suffers from similar electron-beam induced decomposition issues as Mo(CO)\textsubscript{6} inside the source. Under these assumptions, the ionization efficiency of $\mathrm{Mo(CO)_6}$ in a cavity with only cold surfaces (\textit{i.e.} no thermal decomposition), computes to\footnote{The calculation assumes a typical ionization efficiency of 1\% for carbon monoxide, an ionization cross section of ca. \SI{1.9e-20}{\metre\squared} at \SI{120}{\volt} anode potential  \cite{Itikawa2015}, space-charge limited electron current, and a distance between anode and cathode of \SI{1.5}{\milli\metre}.
}
\begin{equation}
\mathcal{E}_{\mathrm{Mo(CO)_6}}^{\mathrm{VD}} \sim \mathcal{E}_{\mathrm{Mo(CO)_6}}^{\mathrm{L}}\; \frac{\mathcal{E}_{\mathrm{Kr}}^{\mathrm{VD}}  \, f_{\mathrm{CO}}^{\mathrm{VD}}}{\mathcal{E}_{\mathrm{Kr}}^{\mathrm{L}}\, f_{\mathrm{Kr}}^{\mathrm{VD}} }\; \sim 2\%.
\end{equation}

The ion source proposed here is exposed to a significant partial pressure of residual carbon monoxide. While in some cases results have been obtained, indicating that residual carbon monoxide might increase the ionization efficiency for elements with lower ionization potential \cite{MartinezPalenzuela:2672954}, experimental results for molybdenum hexacarbonyl are not available. 



A drawback of ion sources exploiting electron impact ionization is the lack of selectivity. Separation of isobaric contaminants in the radioactive ion beam has been achieved by element-dependent adsorption on quartz columns \cite{Bouquerel2007,bouqdiss}. Unfortunately, this technique might not be suitable to separate different carbonyl compounds, as their adsorption enthalpies are in the same range \cite{Even2014}. An alternative approach could be based on differences in compound stabilities. Carbonyl complexes of Tc, Ru and Rh  readily decompose on gold surfaces already at ambient temperature (\SIrange{30}{50}{\celsius}), whereas Mo complexes have a survival probability of \SI{60}{\percent} in the same setup \cite{Usoltsev2017Part1}. Differences in the ionization fragmentation patterns could  be exploited for additional selectivity.

\begin{table*}
	\begin{center}
    \caption{Estimation of expected ion beam yields of \textsuperscript{105}Mo from uranium foils, and \textsuperscript{174}W from platinum foils.}
    \label{tab:totaleff}
    \begin{threeparttable}
	    \begin{tabular}{c|c|c|c|c|c|c|c}
    	Symbol & Unit & Description & \multicolumn{2}{c|}{\textsuperscript{105}Mo}	&	\multicolumn{2}{c}{\textsuperscript{174}W}  & ref. \\
    	\hline
		\rule{0pt}{3ex}  
		$T_{1/2}$ & \si{\second}		& half life				& \multicolumn{2}{c|}{\num{36}}		& \multicolumn{2}{c|}{\num{1860}} & \cite{LundData}\\
    	\hline
		\multicolumn{3}{c|}{Predominant production reaction} & \multicolumn{2}{c|}{U(n,f)} & \multicolumn{2}{c|}{Pt(n,spall)} & sect. 2.1 \\
    	\hline
		\rule{0pt}{3ex}  
    	$N_0$ & \si{\per\micro\coulomb}	& in-target production 	& \multicolumn{2}{c|}{\num{3.8e8}}			&	\multicolumn{2}{c|}{\num{4.5e8}} & sect. \ref{subsec:intarget}\\
    	$t_\mathrm{irr}$ & s 			& irradiation time		& \multicolumn{2}{c|}{50}					&	\multicolumn{2}{c|}{400}	& sect. \ref{sec:concept} \\
    	$I_p$ & \si{\micro\ampere}		& proton current		& \multicolumn{2}{c|}{\num{2.0}}				&  	\multicolumn{2}{c|}{\num{2.0}} & \cite{Catherall2017}\\
   		$N_0^{\mathrm{batch}}$ & 	& isotopes per batch 			& \multicolumn{2}{c|}{\num{2.4e10}}			&	\multicolumn{2}{c|}{\num{3.3e6}} & eq. \ref{eq:yield}  \\
    	$\nu_\mathrm{batch}^{-1}$ & \si{\second} & cycle time	& \multicolumn{2}{c|}{\num{83}}			&	\multicolumn{2}{c|}{\num{454}} & sect. \ref{sec:concept} \\   		
    	\hline
   		\rule{0pt}{3ex}  
    	$\epsilon_\mathrm{extr}$ &	\%	& extraction eff.		& \multicolumn{2}{c|}{10}					& \multicolumn{2}{c|}{1.6}  & sect. \ref{subsec:stopping}\\
    	$\epsilon_\mathrm{stop}$ &	\%	& stopping eff.			& \multicolumn{2}{c|}{49}					& \multicolumn{2}{c|}{21} & sect. \ref{subsec:stopping}\\
	   	$\epsilon_\mathrm{form}$ &	\%	& chemical eff.			& \multicolumn{2}{c|}{80}					& \multicolumn{2}{c|}{30} &  \cite{even2014situ}\\
    	$\epsilon_\mathrm{sep}$	 &	\%	& gas-separation eff.	& \multicolumn{2}{c|}{30}					& \multicolumn{2}{c|}{50} & sect. \ref{subsec:gassep}\\
    		\hline
   		\rule{0pt}{3ex}  
    	$\epsilon_\mathrm{ion}$	 &	\%	& ionization eff.\tnote{$\dagger$}		& 1 & 	0.0015 & 1 & 0.0015 & sect. \ref{subsec:ion}\\
    	\hline
   		\rule{0pt}{3ex}  
    	$N$ 	& \si{\per\second} & average ion rate  			& \num{3.4e4} & 51	& \num{3.9e3} & 6 & eq. \ref{eq:yield} \\

	    \end{tabular}
	    \begin{tablenotes}
	\item[$\dagger$] Two ionization efficiencies are given. The lower efficiency was measured in an exploratory experiment with a cold VADIS, in which electrons were released by a laser. The higher efficiency is an estimated value that is believed to be in reach after development of a cold FEBIAD-type ion source (cf. sect.~\ref{subsec:ion}).
\end{tablenotes}
	  \end{threeparttable}
  \end{center}
\end{table*}

\subsubsection{Laser ionization}

Resonant laser ionization is a powerful tool for element-selective ionization. However, the technique is typically only applied to single atoms. Recently, a resonant laser ionization scheme for molybdenum was developed and tested online at ISOLDE \cite{Chrysalidis:2703661}. Exploiting this scheme for carbonyl compounds requires first to strip the molybdenum atom of its carbonyl ligands. Laser induced breakup of the compound \cite{Leopold_1983,GasPhaseInorgChem} is widely discussed in literature and resonant breakup is also reported in ref.~\cite{windhorn2002molecular}. The concept of laser-induced neutral dissociation followed by resonant laser ionization is further discussed in ref.~\cite{Seiffert:2241995}. While the efficiency of laser induced breakup and ionization has not yet been quantified, the method would allow element-selective ionization. 

\section{Conclusions and Implementation at ISOLDE}
\label{sec:conclusions}

Within the previous sections, we have provided a concept for a gas-filled recoil target, which can be used at ISOL facilities. Thin metallic foils acting as target material are placed around a tungsten rod, which serves as a spallation neutron source.  Instead of diffusion, the recoil energy of the reaction product is exploited to extract the radioisotopes from the foil. They are subsequently thermalized in carbon monoxide gas. Volatile carbonyl complexes form at ambient temperature and pressure. The carbonyl complexes are chromatographically separated from the carbon monoxide gas and fed into an ion source. Following the discussion in the previous section, the estimated efficiency for each step is listed in table~\ref{tab:totaleff}. Experimentally obtained ionization efficiencies in a proof-of-principle setup along with expected values after successful ion source development are given. In the former case, intensities in the order of  \SI{51}{\per\second} for \textsuperscript{105}Mo from uranium foils and \SI{6}{\per\second} for \textsuperscript{174}W from platinum foils can be expected. After successful development of the proposed ion source, a total intensity of \SI{3.4e4}{\per\second} for \textsuperscript{105}Mo and  \SI{3.9e3}{\per\second} for \textsuperscript{174}W is expected. Typically, intensities in this order of magnitude are compatible with post-acceleration within the HIE-ISOLDE linac \cite{Pietro_2017}.\footnote{The actual intensity of the post-accelerated beam depends on additional factors such as presence of isobaric contaminants, molecular break-up and charge state distribution in the electron-beam ion source (charge breeder).}

The implementation at ISOLDE can be achieved with two different approaches. Standard target units already combine a class of target material, some chromatography setup and an ion source \cite{Bouquerel2007}. Following this approach, a compact setup is mandatory that is also suitable to operate in strong radiation fields. The dimensions and weight of the setup are limited by the maximum permitted load for the robot operating the target unit and the design of the target stations. Due to the low temperatures required for gas separation, needs arise to either implement a cryostat system at the target stations, or follow an alternative strategy for gas separation. 

The second approach splits the setup in two assemblies. The gas-filled target remains installed at the target station, while the chromatographic setup is installed remotely, \textit{e.g.} in the ISOLDE experimental hall. The carbonyl compounds are extracted as neutrals in a carbon monoxide stream to the remote location, which is an efficient and established technique, commonly used in transactinide synthesis to transport the compound over a distance of several meters in short times \cite{Even2014}. The split installation circumvents size and weight limitations of the chromatographic setup and the operation of the latter in radiation environments. 

\section{Outlook}

Following this study, we have started to experimentally investigate the feasibility of the concept. A target unit was built, which allows the characterization of the neutron converter setup and the in-target production rates along with beam-induced breakup of carbonyl compounds.
 
The recently started development of a two chamber approach for the synthesis of carbonyl compounds is not yet included in the considerations of this work \cite{Goetz2020}. Within this approach, the target foils can be directly irradiated by the proton beam. The nuclear reaction products are flushed with an inert gas stream into a second chamber, in which the radioactive atoms are allowed to react with carbon monoxide to form carbonyl compounds. In this setup a transport efficiency between the two chambers of more than 50\% was measured. The spatial separation of isotope production and molecule formation would allow benefiting from higher in-target production rates and avoids exposure of delicate compounds to strong radiation fields at the same time.

\section{Acknowledgments}

We would like to thank F. Wenander, U. K\"oster, A. Andreyev and R. Heinke for their comments and fruitful discussions regarding the target concept. We also thank M. G\"otz for the insights into his ongoing development work of the two-chamber approach for carbonyl compound production. We appreciate support for simulation codes from R. Dos Santos Augusto (FLUKA), A. Kelic-Heil and J. Klimo (ABRABLA) and Ch. Duchemin (TALYS and FLUKA). The target unit used in the ionization studies has been manufactured by the ISOLDE workshop (E. Barbero, B. Crepieux, M. Owen,  and A. Vi\'{e}itez Su\'{a}rez). 
This project has received funding from MEDICIS-Promed and the European Union’s Horizon 2020 research and innovation program under grant agreement No 654002.

\section{Authors contributions}
All the authors were involved in the preparation of the manuscript. All authors have read and approved the final manuscript. 
%
\bibliographystyle{epj}

\bibliography{bibtex}
%
%
%

\end{document}